\documentclass[aps, pra, reprint, superscriptaddress, onecolumn, showkeys, longbibliography, notitlepage]{revtex4-1}

\usepackage{graphicx}
\usepackage[english]{babel}
\usepackage[utf8]{inputenc}
\usepackage{amsmath} 
\usepackage{amssymb}
\usepackage{xcolor}
\usepackage{times}
\usepackage{amsfonts,amsmath,amssymb}
\usepackage[margin=35mm,inner=35mm,marginpar=10mm]{geometry}
\usepackage{hyperref}
\hypersetup{
	colorlinks,
	linkcolor={red!50!black},
	citecolor={blue!50!black},
	urlcolor={blue!50!black}
	}

\begin{document}
\noindent

\title{Matter-wave physics with nanoparticles and biomolecules}

\author{Christian Brand}
\affiliation{University of Vienna, Faculty of Physics, Boltzmanngasse 5, A-1090 Vienna, Austria}

\author{Sandra Eibenberger}
\affiliation{University of Vienna, Faculty of Physics, Boltzmanngasse 5, A-1090 Vienna, Austria}
\affiliation{Harvard University, Department of Physics, 17 Oxford Street, Cambridge, MA 02138, USA}

\author{Ugur Sezer}
\affiliation{University of Vienna, Faculty of Physics, Boltzmanngasse 5, A-1090 Vienna, Austria}

\author{Markus Arndt}
\affiliation{University of Vienna, Faculty of Physics, Boltzmanngasse 5, A-1090 Vienna, Austria}

\begin{abstract}
These lecture notes emerged from a contribution to the "Les Houches Summer School, Session CVII--Current Trends in Atomic Physics, July 2016". It is meant to serve students as a guide to a selection of topics that are currently at the focus of molecular quantum optics and complements an earlier lecture on related topics \cite{Arndt2014b}.
In this review we discuss recent advances in molecular quantum optics with large organic molecules. In particular, we present successful experiments of molecules of biological importance, such as neurotransmitters and map out the route towards interferometry of large bio-matter such as proteins. The interaction of internally complex molecules with various beam splitters is discussed for both far-field diffraction at a single nanomechanical grating and the Kapitza-Dirac Talbot-Lau near-field interferometer-- addressing recent progress, experimental challenges, and prospects for quantum enhanced metrology of highly complex systems in the gas phase. A central part of this review deals with the preparation of slow beams of neutral biomolecules, ranging from a single amino acid to proteins or peptides in a micro-environment.
\end{abstract}

\maketitle
\tableofcontents

\newpage

\sloppy

\section{Introduction}

The Les Houches summer school was witness of how far quantum physics with atoms has already been advanced. While the non-classical aspects of quantum physics -- such as mesoscopic superpositions or entanglement -- require us to rethink our notions of reality, locality, logic, or space-time, they have also led to emerging technologies around computation, simulation, metrology, or sensing.   

Our present contribution adds to both lines of research: we discuss how to visualize quantum mechanics in experiments with bodies of increasing mass and complexity, and we demonstrate how quantum interference can enhance molecule metrology and provide a tool for physical chemistry and biomolecular physics. 

Several quantum dualities can be illustrated in such experiments, including the duality of \textit{determinism and randomness}, of \textit{wave and particle}, or the \textit{mutual uncertainty} of conjugate variables. Richard Feynman once pointed out that all these characteristics can be found in a single demonstration \cite{Feynman1965}, the double slit experiment  with massive particles. The discrete, local particle nature appears in the detection process, while the indistinguishability of alternative paths through the experiment -- the delocalized wave nature -- explains the observed interference pattern. This is why Feynman considered the double slit experiment to have "'in it the heart of quantum mechanics"', containing its "'only mystery".	 

Today, we might interject that an important aspect of quantum physics was missing, namely \emph{entanglement} as the fundamental non-separability of two or more quantum systems. However, entanglement is not foreign to matter-wave physics. It appears in the form of mode entanglement and more relevantly in the form of quantum decoherence -- which is even useful as a resource for molecule metrology. 

Double- and multi-slit diffraction experiments with massive matter have been realized with electrons~\cite{Jonsson1961}, neutrons~\cite{Zeilinger1988a}, atoms~\cite{Keith1988,Carnal1991} and their clusters \cite{Schollkopf1994}, as well as small \cite{Schollkopf2004} and large molecules \cite{Arndt1999b}. 
The combination of several diffraction elements into full matter-wave interferometers allowed accessing states of increasing macroscopicity: Nowadays, it is possible to delocalize individual atoms on the half-meter scale \cite{Kovachy2015} and to demonstrate spatial superposition states from single electrons \cite{Hasselbach2010} up to organic molecules exceeding $10^4$~amu~\cite{Eibenberger2013}. 
All studies together already span a factor of $10^7$ in mass and are still fully consistent with Schr\"odinger's quantum mechanics, as developed 90 years ago~\cite{Schroedinger1926}.

In our present lecture we report on explorations of quantum physics with strongly bound, warm objects of high internal complexity. We study matter-wave interference of organic nanomatter that may bind dozens or beyond a thousand atoms into one single quantum object~\cite{Hornberger2012,Arndt2014c}. 
Such experiments aim at pushing quantum physics towards two complementary frontiers: high mass and high complexity; gravity and life.
Both frontiers offer research questions and opportunities for several decades to come. A number of reviews have already been written about the  expected or speculative modifications of matter-wave physics at high mass \cite{Bassi2013b,Arndt2014c}. 
Here, we will focus on the second aspect as a guideline to discuss and develop new techniques for quantum optics with molecules of biological interest. This is driven by five complementary goals: 

\paragraph{Curiosity}
In our daily lives one often tends to make qualitative distinctions between inorganic and organic matter, the inanimate and the animate world or even non-conscious objects and conscious beings. Is there any fundamental limit or is it a distinction of complexity only? Can we still delocalize peptides, proteins, DNA, viruses or cells? This question has triggered a long-term journey, on a mass and time scale comparable with testing the gravitational limits of quantum physics.  

\paragraph{Nanotechnology}
It has been traditionally difficult to prepare neutral beams of size-selected massive particles in the mass range of $10^3-10^6$~amu. Cluster aggregation sources deliver intense beams of particles, but often the signal is distributed over many mass peaks. In contrast to that, life defines and selects the size and shape of biomolecules by their functionality. Can we exploit this 'nature-made nanotechnology' for quantum experiments, even if we do not ask any biological question? Biomolecules are fragile and not easy to volatilize, but progress is being made and some of it is reported here.

\paragraph{Complexity}
Throughout recent years, we have witnessed enormous progress in the handling of few-particle quantum systems and in obtaining control over millions of quantum degenerate ultra-cold atoms, which may be as cold as a few picokelvin. Complementary to that we accept the challenge of exploring quantum physics in a complexity range that is closer to our daily lives, with molecules composed of many hundreds of \textit{covalently} 
bound atoms. Because of their complexity, one can assign an internal temperature to macromolecules, even when they are prepared in spatial Schr\"odinger cat states. We find that the de Broglie wavelength of macromolecules can stay coherent over many milliseconds even when the internal temperatures exceed several hundred Kelvins.

\paragraph{Decoherence}
With increasing complexity decoherence becomes an important agent in the game. The rich internal arrangement, in particular of biomolecules like proteins with their different structure levels -- from primary to even quarternary structure -- opens interaction channels with their environment that do not exist for electrons, neutrons, atoms or dielectric nanoparticles.  

\paragraph{Metrology}
The high sensitivity of macromolecules to external perturbations is both a curse and a blessing: it complicates attempts to prepare pure quantum states but it enables new ways of sensitively measuring molecular properties -- by monitoring their matter-wave fringe shifts in the presence of external forces. 
Such measurements build on the same principles as modern atom gravimeters and rotation sensors. While molecule interferometers will not compete with atom sensors as inertial sensors, their unique potential is rooted in their capability to act universally on a variety of particle classes and to expose them to external fields -- as a unique test bed for physical chemistry and biomolecular physics, with the potential of becoming a viable technology.

In these lecture notes, we will present a tutorial introduction to molecular quantum physics, focusing on the coherent state preparation, on the diversity of diffraction techniques for complex molecules, on methods how to exploit quantum interference for obtaining molecular spectra and examples of challenges in preparing and detecting neutral biomolecular beams for quantum experiments.

\section{Delocalization and diffraction}

\subsection{General source and coherence requirements}
In this lecture we are focusing on quantum aspects of the center-of-mass motion of complex molecules, i.e.\ on the coherent splitting of their de Broglie wave fronts, their recombination and interference. In our experiments, de Broglie wavelengths range between $\lambda_\mathrm{dB}=h/mv=0.3-5\times 10^{-12}$~m, which is typically $10^3-10^4$ times smaller than the size of the molecule itself and comparable to the shortest wavelengths in high-resolution transmission electron microscopy. For particles lighter than about 1000~amu, it is often possible to sublimate or evaporate them in thermal Knudsen cells. This even holds for a sizable set of biomolecules such as nucleobases, vitamins, antibiotics, vanillin, caffeine and various other compounds. Thermal beams have therefore proven useful for many of our quantum interference studies so far. 

In molecular beams of small divergence -- even in near-field interferometers the molecular beam is typically constrained to a divergence angle of 1~mrad -- we can distinguish longitudinal and transverse coherence, which provide  a measure for the interference capability of the emerging molecules. It is determined by the spectral and geometric properties of the source.

\textit{Longitudinal coherence, } as in light optics, is determined by the spectral purity of the source. A practical value is  $L_c=\lambda^2_\mathrm{dB}/\Delta \lambda_\mathrm{dB}$, defined by the relative spread $\Delta \lambda_\mathrm{dB}$ of de Broglie wavelengths $\lambda_\mathrm{dB}$. It remains constant with distance from the source but can be improved by spectral filtering, i.e.\ velocity selection. In thermal sources this value can be as short as $L_c\simeq 2\lambda_\mathrm{dB}$, still much smaller than the molecule itself. Again, this does not prevent successful interference since we are only dealing with the description of the center-of-mass motion of a single particle.

\textit{Transverse coherence }is at the origin of the indistinguishability of different molecular paths through space. It can be estimated both from Heisenberg's uncertainty principle in quantum physics and the van Cittert-Zernike theorem in optics. Transverse coherence $X_c\simeq 2 L \lambda_\mathrm{dB}/D $  grows linearly with the distance from the source $L$ and inversely proportional with the source size $D$. Being based on a diffraction phenomenon, it increases with increasing de Broglie wavelength.
	
While path-indistinguishability is a key condition for matter-wave interference, we may also ask whether there is any chance to establish two-particle interference, as in Hanbury-Brown-Twiss experiments, which have already been realized for atoms~\cite{Westbrook1998}. This would require identity in all degrees of freedom. Any $N$-atom system has $3N$ mechanical degrees of freedom, in addition to electronic, fine, hyperfine and Zeeman levels in a complex configuration space. Large molecules have a conformational orientation, permanent and vibrationally-induced dipole moments, and may undergo conformation changes upon photo-absorption (photo-isomerization). 

This is why macromolecule interferometry deals only with the superposition of single-particle wave functions. Molecule lasers, in the spirit of successfully realized atom lasers~\cite{Bloch1999,Andrews1997,Hagley2001}, would require sub-microkelvin-low temperatures in all degrees of freedom and even then a mechanism to ensure that all molecules end up in the same absolute energy minimum of the conformational landscape. While this feat has been achieved for diatomic molecules -- it remains an outstanding challenge for macromolecules.

In thermal sources the mechanical degrees of freedom are equilibrated with the source temperature in the range of $500-1000$~K by collisions. However, as soon as the molecules leave the cell and enter high- or even ultra-high vacuum, these degrees of freedom are decoupled on the time scale of the interference experiments. In the absence of thermal emission processes, the micro-canonical temperature then determines the rotation and vibration of the molecule -- and the exchange of energy between the vibrational modes on a picosecond time scale -- but it remains decoupled from the center-of-mass-motion.

Since matter-wave physics requires only a well-defined center-of-mass state of an ensemble of molecules whose internal states are sufficiently similar, de Broglie interference can actually be observed for surprisingly large composite objects.

\subsection{Beam splitting of molecular matter-waves}
Coherent beam splitters are the foundation of every interferometer. They divide, redirect and recombine the incident waves. In analogy to classical optics, we distinguish two species also in matter-wave interferometry: 

\textit{Wave front beam splitters} modulate the incident wave front using a spatially periodic structure, i.e.\ a diffraction grating, which may either modulate the matter-wave phase or amplitude. \textit{Phase gratings} for atoms or molecules can be realized by exploiting the dipole potential emerging from the interaction between the particle's polarizability and the electric field amplitude of a standing light-wave grating. \textit{Absorptive gratings} may for instance be built using nanomechanical masks or \textit{photo-depletion} mechanism such as single-photon-ionization, 2-photon ionization, or photo-fragmentation in the anti-node of a standing light field.

\textit{Amplitude beam splitters}, such as Ramsey-Bord\'e~\cite{Borde1989} or Raman beam splitters~\cite{Kasevich1991} do not realize stationary gratings in real space,  but create entanglement between the internal atomic state and the external center-of-mass motion during the coherent interaction between the atom and one or two laser beams. We will see further below, that even in the absence of any coherent cycling as in atoms, single-photon absorption can contribute to the beam splitting process of hot molecules~\cite{Cotter2015}. 

Here we review the concepts of beam splitters for complex molecules and discuss the dephasing mechanisms that emerge in particular in experiments with biomolecules, which are typically polar.

\subsection{Far-field diffraction at a nanomechanical grating}\label{sec:FarField}

First demonstrations of matter-wave physics with electrons~\cite{Davisson1927c}, atoms, and diatomic molecules~\cite{Estermann1930} were done using reflective diffraction at clean crystal surfaces. With the advent of nanotechnologies, free-standing nanomechanical transmission gratings were realized for electrons~\cite{Jonsson1961}, atoms~\cite{Keith1988}, and molecules~\cite{Schollkopf2004,Arndt1999b}. Such beam splitters are open masks that remove portions of the incident wave front by absorption or scattering. The spatially periodic confinement of the wave function in the slit openings leads to a spread of the transversal momentum behind the mask, in accordance with Heisenberg's uncertainty principle and diffraction rules as known from wave optics. In contrast to light optics, the matter-wave transmission function $\tilde{t}(x)$ is, however, not described by a binary filter $t(x)$ alone, but also by the attractive interaction between the particles and the grating walls.

In the simplest case, we can treat the local interactions by considering particles that pass the grating wall in a straight line at distance $x$. In the Eikonal approximation we assume that this trajectory receives a transverse momentum kick which is soft enough that the related position shift will only become noticeable long after the grating. The interaction time is determined by the grating thickness $b$ and the molecular velocity $v_z$ along the forward direction. For thick gratings with openings as narrow as 50~nm it is also fair to assume that we can neglect retardation effects and treat the Casimir-Polder (CP) interaction in the near-field approximation, i.e.\ using a van der Waals scaling~\cite{Grisenti1999,Brand2015c}:

\begin{equation}
\tilde{t}(x)=t(x) \, \exp\left(\frac{i}{\hbar v_z}\int_{0}^{b}{\frac{C_3}{x^3}dz}\right)
\label{eqn:mechanicaltransmission}
\end{equation}

The phase is determined by the molecular polarizability $\alpha(\omega)$ at all frequencies $\omega$ as well as by the dielectric function $\varepsilon(\omega)$ of the grating material, which is encoded in the surface reflection coefficient $r(\omega)=(\varepsilon(\omega)-1)/(\varepsilon(\omega)+1)$.

\begin{equation}
C_3=\frac{\hbar}{16\pi^2\epsilon_0}\int_0^\omega{\alpha(\omega)r(\omega)d\omega}
\end{equation} 

Analysis of atomic diffraction patterns have yielded detailed information about atomic polarizabilities~\cite{Holmgren2010} and atom-surface interactions~\cite{Perreault2005b,Grisenti1999}.

The description gets more demanding for molecules because of their complex internal structure and dynamics. They may have a permanent or even vibration-induced electric dipole moment and they may appear in a plethora of different conformational arrangements. Hence, the attractive interaction to the wall depends on the molecular conformation and orientation inside the slit. At a source temperature of 500-1\,000~K the molecules are highly excited and conformations can interconvert on the time scale of 1-10~picoseconds.  This has to be compared to the molecular transit time through the grating, which amounts to 500~ps for $v=200$~m/s and $b=100$~nm. 

Hence, some the molecule-wall interaction can be described by an average value over molecular starting conditions and orientations. On the other hand, additional effects become relevant. Vibrationally induced dipole moments, for instance, will increase the effective static polarizability~\cite{Gring2010b} and the molecule may exchange excitations with the grating during the transit, leading to additional attractive or repulsive contributions to the potential~\cite{Brand2015c}. 
\begin{figure}[htb]
\centering
\includegraphics[width=0.9\linewidth]{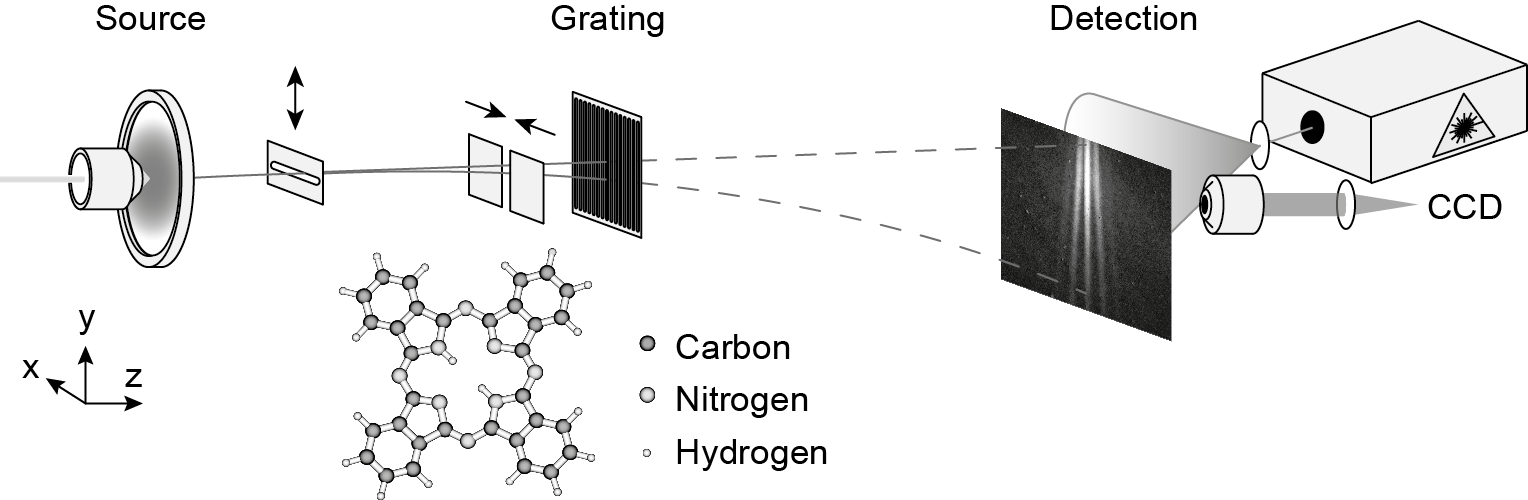}
\caption{Molecules like phthalocyanine shown here are thermally evaporated by a continuous laser beam (421~nm) focused down to a spot size of $\sim$1.5~$\mu$m. The resulting molecular beam is shaped in $x$- and $y$-direction before it reaches the grating after 1.55~m. The molecules illuminate 50-90 slits coherently. After additional 0.59~m they arrive at the quartz window of the detector. The quantum nature of the diffraction is revealed by analyzing the position distribution of the arriving particles, which we image using laser-induced fluorescence microscopy.}
\label{fig:Farfield}
\end{figure}
The current setup of the far-field diffraction apparatus is shown in Fig.~\ref{fig:Farfield}~\cite{Juffmann2012}. It is housed in a 2.14~m long vacuum chamber which is evacuated to a pressure below $10^{-7}$~mbar. This is necessary to guarantee a collision-free flight of the molecules through the whole apparatus. The journey of the molecules begins in the source. They are deposited as a thin film on the inner surface of a vacuum window. When we focus a laser through the glass onto this film, we can evaporate the molecules thermally, which requires that the laser is resonant to an electronic transition of the molecule. The size of the laser spot defines the source size $D$ and is chosen as small as possible. For a spot size of $1-2~\mu$m the transversal coherence of a particle with a wavelength of $\lambda_{dB}=3$~pm amounts to $2L\lambda_{dB}/D=5-9~\mu$m at a distance $L=1.55$~m behind the source. For common gratings with a period of 100~nm this corresponds to a coherent illumination of 50 to 90 slits. After diffraction the molecules propagate for another 59~cm before they impinge on a thin quartz plate where they stick. In order to visualize the molecular pattern we need a method to read out the position of each particle with high accuracy. This is done by exciting the molecules electronically with laser light of suitable wavelength. Once elevated to the excited state the molecules release the excess energy by the emission of a photon. For systems with a high fluorescence quantum yield this can be done many times before the molecule undergoes a photochemical reaction and is lost to the detection process. Collecting a large number of photons allows for a very precise localization of every molecule down to 10~nm~\cite{Juffmann2012}. 

\begin{figure}[thb]
	\centering
	\includegraphics[width=0.9\linewidth]{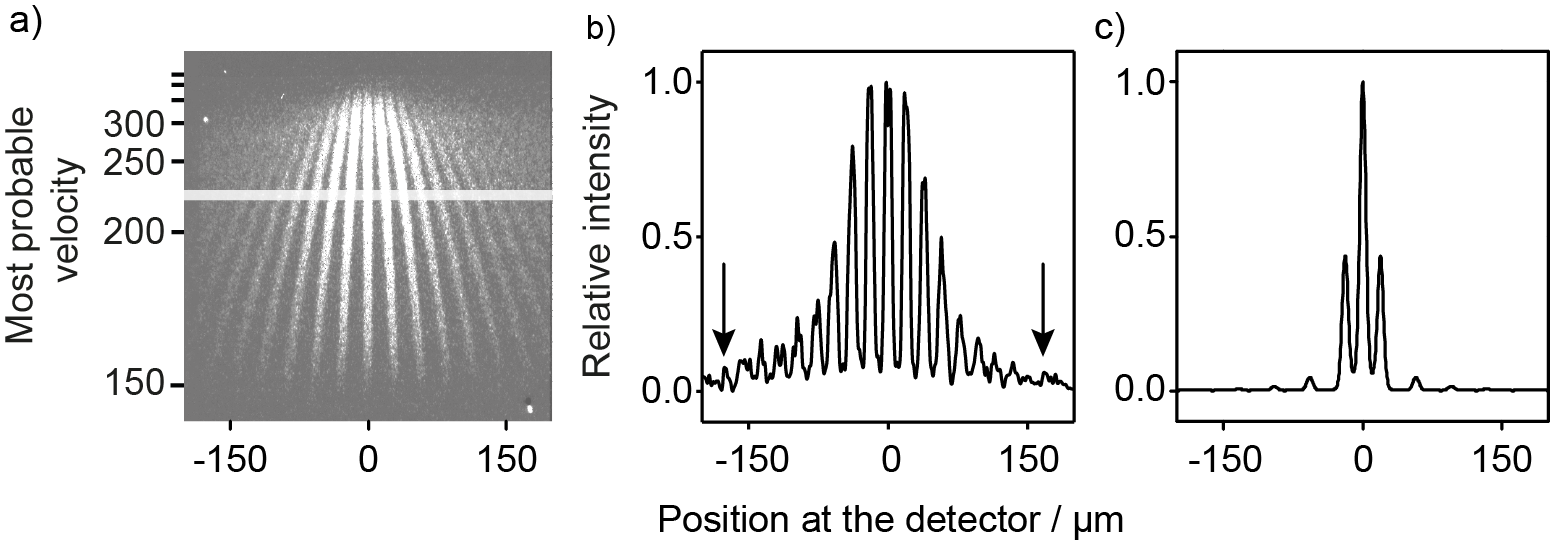}
	\caption{a) Molecular diffraction of phthalocyanine at a 45~nm thick silicon nitride grating. The different molecular velocities are sorted by gravity and range between 350 and 150~m/s. The trace in (b) corresponds to the highlighted area in (a) and the arrows point at the $\pm 9^{th}$ diffraction order. The population of high diffraction orders can be traced back to the Casimir-Polder interactions. The huge impact of these can be estimated when we compare the experimentally observed trace to a simulated one neglecting any attractive interactions (c).}
	\label{fig:SiN_diffraction}
\end{figure}

A typical diffraction pattern is shown in Fig.~\ref{fig:SiN_diffraction}. It shows the signal of the organic dye phthalocyanine PcH$_2$ diffracted at a 45~nm thick membrane made of silicon nitride \cite{Brand2015a}. Into this our collaborators around Ori Cheshnovsky at Tel Aviv University milled a mask using a focused gallium ion beam. We can extract various information from this image. First, we see high-contrast diffraction, up to  the $9^{th}$ order. This proves the de Broglie wave nature of PcH$_2$ molecules and corroborates the assumption that quantum mechanics in the translational degrees of freedom can persist, even if the internal degrees of freedom are highly excited. It also underlines that the rotational symmetry, which we assume for atoms and could also take for granted in earlier C$_{60}$ diffraction experiments~\cite{Arndt1999b} is no prerequisite for the preparation of a clean translational quantum state.
The spacing of the diffraction orders increases from the top to the bottom of the image. Since slower molecules fall further down in the gravitational potential of the earth before they reach the detector, their height encodes the molecular speed and de Broglie wavelength. The highest diffraction orders on both sides of the image are separated by $\pm9$ grating momenta. For comparison, we calculate how many momenta of light would be necessary to realize the same separation for rubidium in a light grating at 780~nm. This yields

\begin{equation}
\Delta p= 18 \cdot \frac{h}{d}=\frac{18}{100\,\mathrm{nm}} \cdot \frac{h}{2\pi}\cdot \frac{2\pi}{780\,\mathrm{nm}}\cdot  \,780\,\mathrm{nm}=140\,\hbar k_{Rb}, 
\label{eqn:LMBS}
\end{equation}

which compares very favorably with large momentum transfer beam splitters in state-of-the-art atom interferometry~\cite{Muller2008a,Kovachy2015}.

The enhanced polarizability of complex molecules may thus actually be useful for achieving a wide arm separation in nanomechanical beam splitters, since the population of high order diffraction orders is enhanced by the van der Waals interaction. Future efforts will still have to focus on improvements of the grating's coherence, i.e.\ the regularity of the grating period to about 10~ppm over a few millimeters in  a single mask as well as between different masks of the same fabrication batch. For 160~nm thick gratings written photo-lithographically this is already proven state of the art for more than two decades~\cite{Savas1995}. Modern focused ion beam (FIB) machines with interferometrically controlled translation stages are approaching a similar precision. A technological exploitation of such large momentum beam splitters requires either or blazed gratings or more intense molecular sources, since the signal is still distributed over many diffraction orders. 

A proper description of diffraction patterns also includes details of the grating bars, such as their geometry and wedge angles~\cite{Savas1995,Grisenti2000b} and material inhomogeneities. The model can be based on a perturbative approach, where it is often sufficient to include only the scattering of a virtual photon from the molecule to the grating, and back. Even scattering inside the material can be corrected for~\cite{Brand2015c}. Any remaining discrepancy between theory and experiment provides a hint on additional interactions, such as charges inside the thin dielectric membrane. Our experiments with PcH$_2$ diffracted at a FIB written grating showed an attractive potential 5-8 times stronger than expected based on the CP-interaction alone~\cite{Brand2015c}. This effect was greatly reduced for masks that were generated in photo-lithography and reactive ion etching at low energies. 

For non-polar molecules the near-field CP-force and the attraction between a neutral particle and a residual charge have the same distance scaling and can be hardly distinguished. However, for polar molecules the interaction with the grating depends on the orientation of the molecular rotation axis to the wall and to the electric dipole moment. This has clear consequences.
\begin{figure}[thb]
\centering
\includegraphics[width=0.7\linewidth]{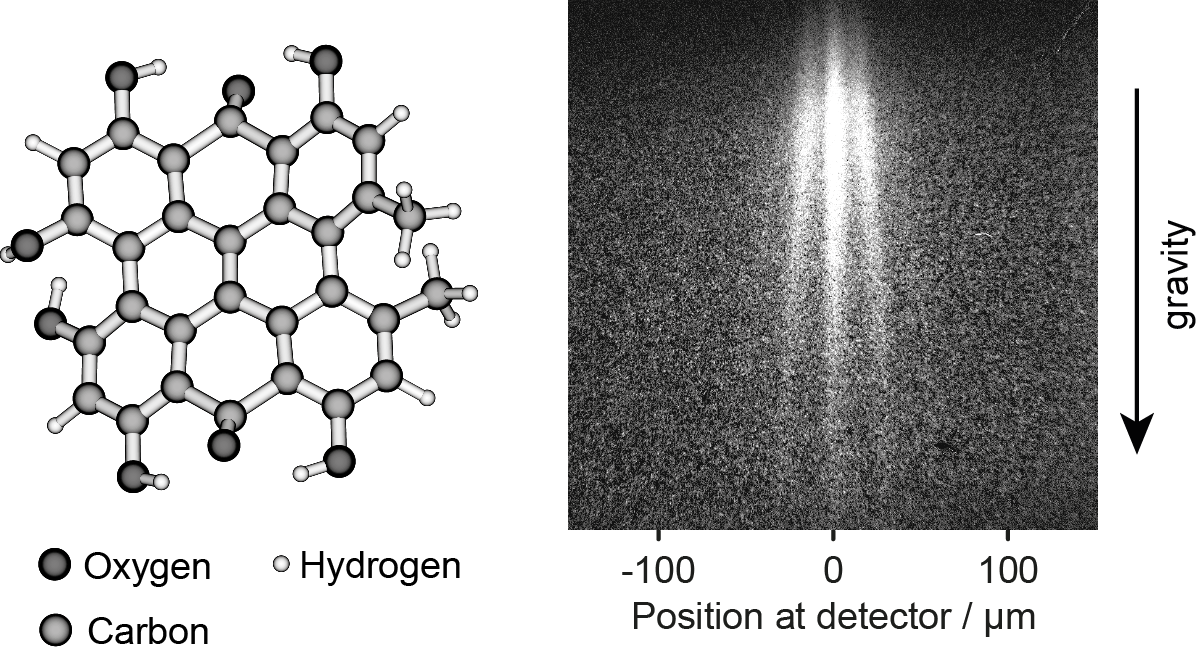}
\caption{Even the polar neurotransmitter and antibiotic hypericin still shows interference, when diffracted at a nanomechanical carbon grating of 20~nm tickness and 100~nm period. However, its permanent dipole moment is too large to allow diffraction at an insulator mask such as SiO$_2$. }
\label{fig:Hypericin}
\end{figure}
It is of particular importance for the diffraction of biomolecules at nanomechanical masks. They are all equipped with functional groups such as OH, NH$_2$ or COOH, which determine the interaction with their surroundings and are necessary for their biological function. Hypericin, naturally occurring in the medicinal herb Saint John's wort, acts as a neurotransmitter, antibiotic, and antiviral agent~\cite{Knobloch2016a}. It serves here as an example of a polar biomolecule. Figure~\ref{fig:Hypericin} shows its  diffraction pattern at a carbon grating of 20~nm thickness. The diffraction peaks are less sharp than those of phthalocyanine in Fig.~\ref{fig:SiN_diffraction}, which we attribute to the permanent electric dipole moment of hypericin of about 4~Debye. To elucidate the role of the molecular polarity, we have compared the diffraction pattern of the non-polar tetraphenylporphyrin (TPP) and its polar derivative (MeO)TPP. Both substances are close in mass and polarizability, but differ in their dipole moment because of the OCOCH$_3$ group attached to one phenyl ring. 
\begin{figure}[htb]
\centering
\includegraphics[width=0.9\linewidth]{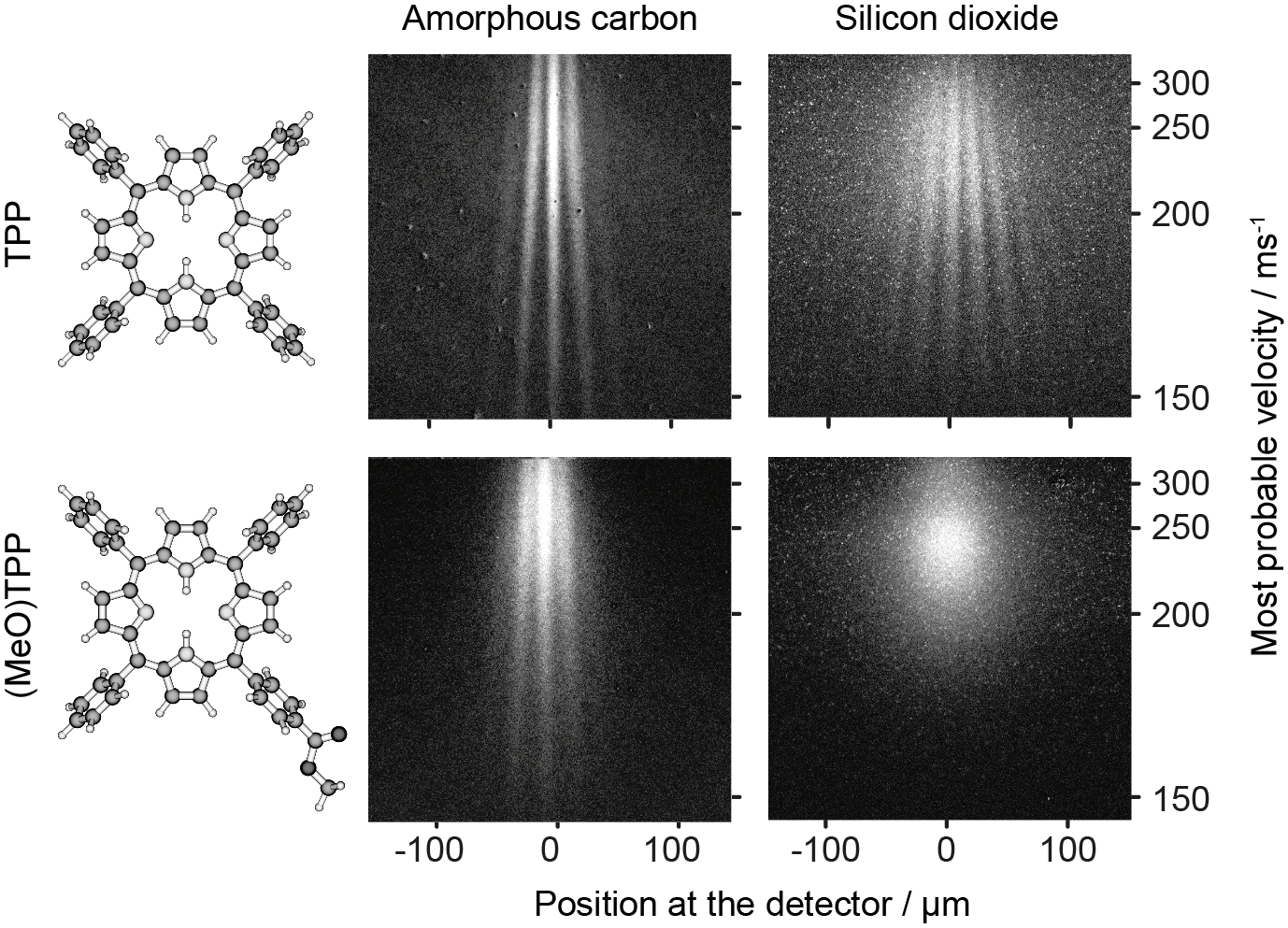}
\caption{Molecular diffraction patterns of the non-polar TPP (C$_{44}$H$_{30}$N$_4$, $m=615$~amu) and the polar (MeO)TPP (C$_{46}$H$_{32}$N$_4$O$_2$, $ m=673$~amu, dipole moment 4~Debye) at mechanical gratings written in amorphous carbon (period d=100~nm; left) and silicon dioxide (d=160~nm; right). The resistivity of amorphous carbon ($1.55 \times 10^{-5}~\Omega$~m) is 20 orders of magnitude smaller than of silicon dioxide ($10^{15}~\Omega$~m). This readily explains why a carbon grating can be substantially less or more homogeneously charged and better suited for diffracting of polar molecules.
}
\label{fig:TPP-MeOTPP}
\end{figure}
In Fig.~\ref{fig:TPP-MeOTPP} their diffraction patterns are compared for two different gratings~\cite{Knobloch2016a}, both milled by a focused ion beam, but into materials with greatly differing conductivity. While the carbon membrane is a weak electric conductor the silicon dioxide grating is an insulator. Hence, charges buried inside the gratings are more likely to be neutralized inside the carbon layer than in  silicon dioxide. 

The non-polar TPP shows high contrast interference at both materials.
The polar (MeO)TPP, however, exhibits already a slightly blurry pattern behind the carbon grating and it is entirely washed out behind the silicon dioxide mask. Not a single diffraction order can be resolved behind the insulating grating and slow molecules even seem to be deflected to beyond the detector area.  

This observation is attributed to the dynamics of the permanent electric dipole moment in an inhomogeneous electric field. In contrast to non-polar molecules, where the interaction is always attractive, polar molecules may experience attractive or repulsive forces depending on their orientation with respect to local charges which may implanted during focused ion beam writing in SiO$_2$ with a density up to $10^{12}$/cm$^2$~\cite{Yogev2008}. 

In thermal beams, neither the initial orientation nor the rotational state are well controlled. Furthermore, the electric field inside the dielectric can be stochastically distributed in amplitude and direction. In consequence, each molecule will experience a different local force and we accumulate the incoherent sum of many coherent but shifted single-molecule diffraction patterns on the screen. Even though there is no proper decoherence, path entanglement, or which-path information shared with the environment, the matter-wave contrast disappears because of this random phase averaging, due to the molecule-wall interaction. This is of particular importance for large polar biomolecules, such as for instance insulin which is known to have an electric dipole moment of 72~Debye in solution~\cite{Laogun1984}.  

However, even for non-polar molecules the CP-interaction influences the molecular diffraction pattern substantially, as shown in Fig.~\ref{fig:SiN_diffraction}. The simplest approach to modeling this pattern is to encode the attractive interaction in an 'effective width' of the single slit diffraction pattern, which defines the fringe envelope~\cite{Grisenti1999}. Fitting this width allows extracting a qualitative measure for the path-integrated interaction strength. During the diffraction of PcH$_2$ at a 45~nm thick SiN$_\mathrm{x}$ grating, for instance, the geometrical slit width of 50~nm appears reduced to an effective width of only 15~nm~\cite{Brand2015a}. Larger molecules with a higher polarizability will see a further reduced slit width and may eventually not even pass the grating any more. 

\begin{figure}[htb]
\centering
\includegraphics[width=0.8\linewidth]{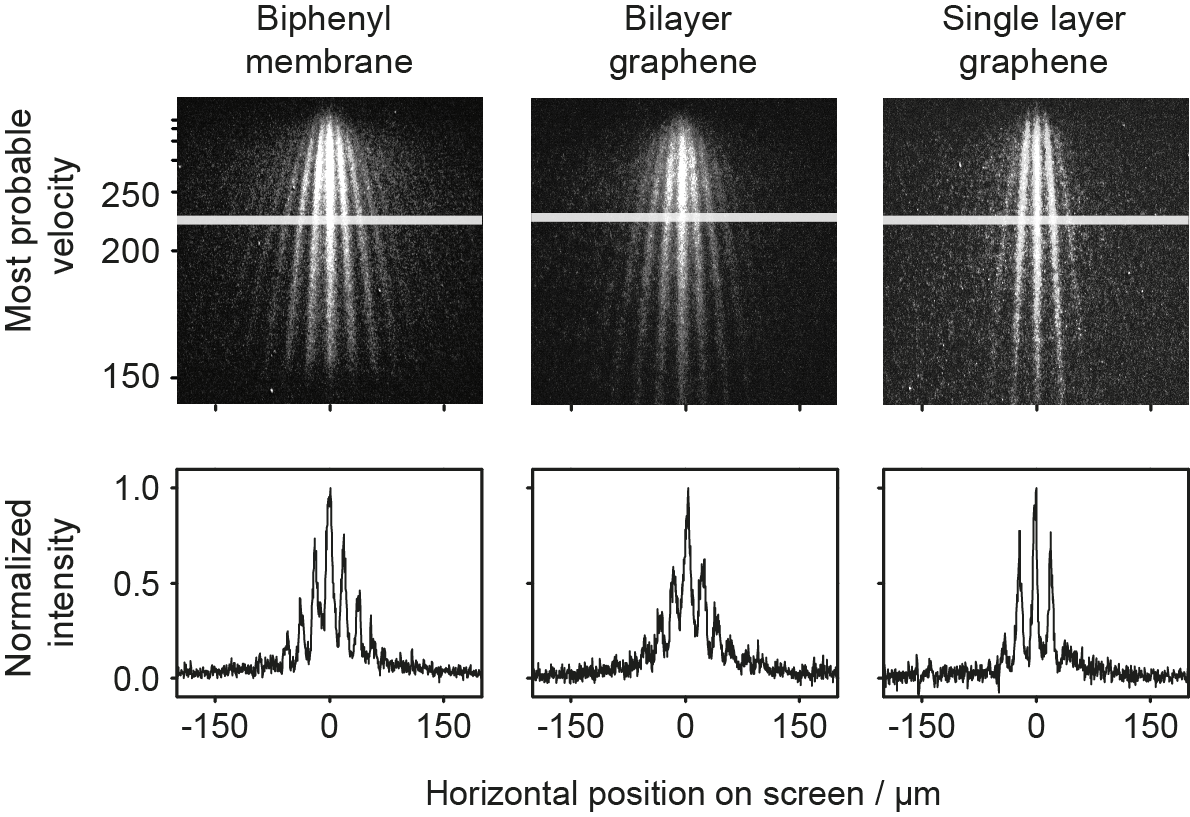}
\caption{Molecular diffraction at gratings with a membrane thickness of 1~nm. While the biphenyl membrane (left) is an insulator, bilayer (middle) and single-layer (right) graphene are conducting. The reduced population of higher diffraction orders is due to the  reduction of the molecule-wall interaction.}
\label{fig:Ultrathindiffraction}
\end{figure}

For quantum experiments with massive particles it therefore appears essential to reduce the grating thickness -- and thus the relevance of CP-potentials -- to the thinnest conceivable value. In recent years, 2D-membranes have become available, that range in thickness between a few nanometers and a single atomic layer. Our collaboration partners at Tel Aviv University made it possible to manufacture gratings with 100~nm period into such membranes with high quality~\cite{Brand2015a}. In Fig.~\ref{fig:Ultrathindiffraction} we show the diffraction patterns of PcH$_2$ behind several of these gratings. They all are about 1~nm thick and we observe a clear reduction of the CP-force compared to thicker masks. The smallest interaction is indeed observed for single layer graphene, which is only one carbon atom thin. However, even this ultimate nanomechanical element for molecular quantum optics exhibits an effective slit width of 35~nm, still 40 percent smaller than the geometrical opening. Although the molecule spends only a few picoseconds 'inside' the slit -- the molecule is actually larger than the grating thickness (!) --  it is still affected by the CP-interaction. 
These experiments also allude to a discussion between Nils Bohr and Albert Einstein, on whether information about the path of a particle through a double slit can be retrieved by monitoring the recoil of the mask. If this were possible, this would destroy the diffraction pattern~\cite{Bohr1949a}. Single layer graphene is about the lightest and thinnest durable mechanical mask one can think of, making it a valid candidate for testing this question. According to Bohr the diffraction remains unnoticed and coherence maintained when the momentum transfer between molecule and mask is smaller than the intrinsic momentum uncertainty of the grating itself. 
The following back-of-the envelope estimate may elucidate this relation: The momentum uncertainty for the matter-wave diffracted at a rectangular slit is~\cite{Nairz2003}:
$\Delta x_{\mathrm{mol}} \cdot \Delta p_{\mathrm{mol}}\geq 0.89 \,h$.
This relation looks different from the one typically shown in textbooks, since we define $\Delta p_{\mathrm{mol}}$ as the experimentally observable full width at half maximum (FWHM) of the diffraction envelope and $\Delta x_{\mathrm{mol}}$ as geometrical width of a single diffraction slit. For the grating we take $\Delta x_{\mathrm{grat}} \cdot \Delta p_{\mathrm{grat}}\geq \hbar$, where the uncertainties are usually taken to be the 1/$\sqrt{e}$ values of a Gaussian distribution in position and momentum. The threshold for coherent diffraction is $\Delta p_{\mathrm{grat}}=\Delta p_{\mathrm{mol}}$. As soon as $\Delta p_{\mathrm{mol}}$ gets larger than $\Delta p_{\mathrm{grat}}$ coherent diffraction vanishes. When we take the minimal uncertainty in momentum of the grating $\Delta p_{\mathrm{grat}}=\hbar/\Delta x_{\mathrm{grat}}$ and insert this into the other equation, we get

\begin{equation}
\Delta x_{\mathrm{mol}}\geq \frac{0.89\, h}{\Delta p_{\mathrm{mol}}}\Rightarrow\Delta x_{\mathrm{mol}}\geq \frac{0.89\, h\cdot \Delta x_{\mathrm{grat}}}{\hbar}\Rightarrow\Delta x_{\mathrm{mol}}\geq 1.78\pi \Delta x_{\mathrm{grat}} .
\end{equation}

This is the lower boundary for the geometrical opening of the grating at which diffraction is still coherent. For graphene nanoribbons the measured position uncertainty amounts to $\Delta x_{\mathrm{grat}}=0.1-0.5$~nm~\cite{Sapmaz2003,Garcia-Sanchez2008}. Hence, each slit would have to be smaller than about 3~nm. In this limit, van der Waals forces would extract the molecules from the beam, anyhow.

\subsection{Near-field interferometry using optical gratings}
\subsubsection{Phase gratings}
In contrast to material gratings, which inevitably introduce an orientation-dependent phase shift, optical beam splitters can act on a large variety of particles, even when they are charged or exhibit an electric or magnetic dipole moment. 
The internal complexity of large molecules renders coherent cycling between the ground and the excited states difficult and excludes Ramsey-Bord\'e or Raman interferometers. We can, however, exploit all optical processes that induce a spatially periodic phase shift or entail a spatially periodic depletion of the molecular beam. 

The diffraction at a thin standing light wave has entered the literature already in 1933, when Piotr Kapitza and Paul Dirac suggested electron reflection at an optical Bragg grating~\cite{Kapitza1933}. The original Kapitza-Dirac (KD) effect is based on the ponderomotive potential that the electron experiences in the presence of a light beam~\cite{Batelaan2007}. The electron itself has no known internal structure but charge: it follows the electric field and in motion it sees the Lorentz force of the magnetic field component, too. The first demonstration of KD-diffraction with electrons was achieved in 2001~\cite{Freimund2001}.

Phase gratings for atoms were already realized fifteen years earlier than for electrons~\cite{Gould1986} and all-optical Mach-Zehnder atom interferometers were demonstrated with thin~\cite{Rasel1995,Delhuille2001} and thick phase masks~\cite{Giltner1995a}. Here, the spatially periodic phase imprint is based on the dipole interaction between the electric field of the standing laser wave and the particle's optical polarizability~$\alpha_{opt}$. This interaction induces an electric dipole moment oscillating at the laser frequency $\omega_L$, far-off any molecular resonance. The coupling of the dipole moment to the electric field $E$ in the standing wave results in the potential $V(x,z,t)=-\alpha_{opt} |E(x,z,t)|^2/4$. 

The spatially periodic transmission function $t(x)=\exp(i\phi(x))$ can be calculated from the phase $\phi(x)$ accumulated during the molecular transit across the standing laser light wave of power $P_L$ and period $d=\lambda_\mathrm{L}/2$ along the $x$-axis (transverse to the molecular beam) with a Gaussian beam waists $w_y$ along the $y$-axis and $w_z$ along the $z$-axis (longitudinal)~\cite{Arndt2014b}.
For a pure phase grating, the accumulated phase $\phi(x)$ is given by

\begin{equation}
\phi(x)=\phi_0 \cdot \cos^2(k_L x) = \frac{4\sqrt{2\pi} \alpha_{opt}P_L}{hc\varepsilon_0 w_y v_z}\cdot \cos^2(k_L x),
\label{eqn:phaseperiod}
\end{equation}

For eqn.~\ref{eqn:phaseperiod} we have assumed that the molecular absorption cross section is sufficiently small that we can neglect any uptake of a real photon by the molecule.

The far-field diffraction pattern at the sinusoidal phase grating is determined by the Fourier transform of the transmission function and proportional to a sum over Bessel functions $J_n$,

\begin{equation}
\psi(x) \propto \sum_{n=-\infty}^{\infty} J_n(\phi_0) \exp \left(-i n 2\hbar k_\mathrm{L} x\right), \mbox{with\,\,} n\in \mathrm{N}.
\label{eqn:phaseshift}
\end{equation}

We learn from eqn.~\ref{eqn:phaseshift} that the momentum exchange between the molecule and the light wave always involves a multiple of two photon momenta, $2\hbar k_\mathrm{L}$. This can be interpreted as the absorption and stimulated emission of virtual photons, but the derivation does not require the photon picture at all. It suffices to consider the spatial geometry of the potential landscape. This is also the reasons why the de Broglie wavelength does not enter in this consideration. For a structure of period $d$ the momentum transfer always amounts to multiples of $h/d$.

Far-field diffraction experiments with fullerenes~\cite{Nairz2001b} showed that this idea can also be used for large molecules, in spite of their internal complexity, high thermal excitation and transition lines as wide as 20-40~nm~\cite{Dresselhaus1998}. Even though textbooks always display grating diffraction with an interference maximum at the center of the pattern, eqn.~\ref{eqn:phaseshift} explains why this is no longer the case for phase gratings: Only at low laser power or small molecular polarizability the sum over the Bessel functions will mimic the far-field diffraction pattern of an absorptive grating, like the graphene nanomasks in section~\ref{sec:FarField}. At high laser power, the central peak can be suppressed.

\subsubsection{Kapitza-Dirac-Talbot Lau interferometry in the near-field}
Beam splitters based on wave front division in a phase grating are an interesting option for composite particles, but the observation of wave interference in the far-field requires the collimation angle to be smaller than the diffraction angle. With decreasing de Broglie wavelength, we therefore need to collimate the molecular beam increasingly well. This becomes demanding for particles the size of a protein. 

A solution to this problem is near-field interferometry which allows working with spatially incoherent and wide particle beams. This results in a huge gain in signal. However, it comes at the price that we now need two or more diffraction elements to prepare and probe the molecular de Broglie wave nature. The idea of using Talbot-Lau interferometry for that purpose goes back to John Clauser~\cite{Clauser1997} who also demonstrated it for thermal potassium beams~\cite{Clauser1994a}.

The idea was successfully implemented at the University of Vienna for fullerenes~\cite{Brezger2002a}, porphyrins, and even fluorinated C$_{60}$F$_{48}$~\cite{Hackermuller2003a} -- using large micromechanical gold gratings with 500~nm thickness and grating periods of $d=990$~nm.
\begin{figure} [tbh]
	\centering
	\includegraphics[width=0.9\linewidth]{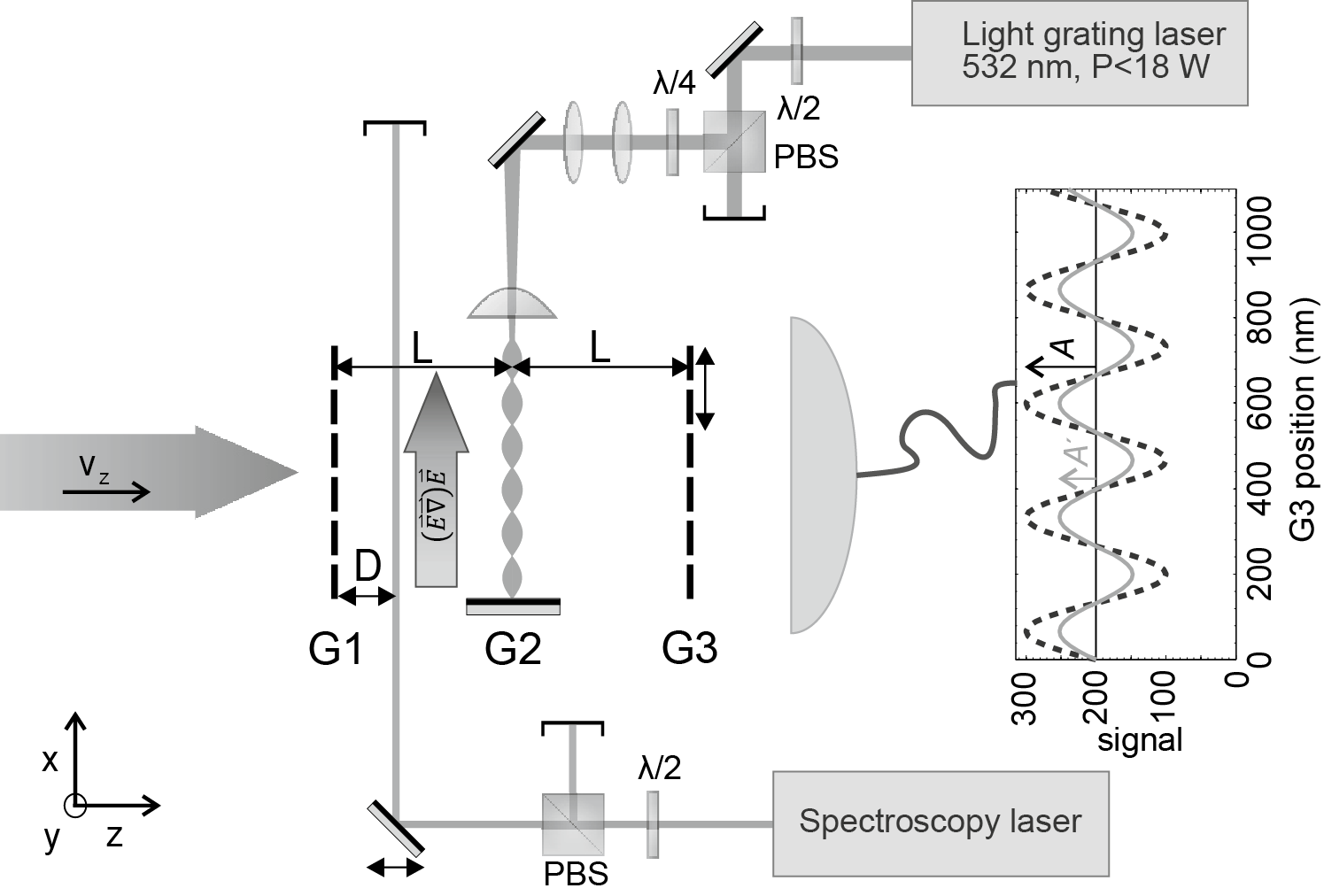}
	\caption{Kapitza-Dirac-Talbot-Lau interferometer (KDTLI): Molecular coherence is prepared by diffraction at each slit of the first grating G1. The coherence function spreads out and covers more than two anti-nodes of the optical grating G2. The spatially periodic phase imprinted by the standing light wave and subsequent interference lead to the formation of a molecular density pattern at the third grating, G3. It serves to mask the molecular fringe pattern, before the molecules are ionized and counted in the quadrupole mass spectrometer. An intense spectroscopy laser and high-voltage deflection electrodes can interact with the molecules to shift their interference pattern and to retrieve information about their internal molecular properties.}
	\label{fig:kdtli}
\end{figure}
However, it soon became clear that the dispersive forces of massive organic molecules near mechanical walls would render the source requirements exigent, in particular with regard to velocity selectivity~\cite{Brezger2003c}. The solution to this challenge was to combine an absorptive grating with a phase grating and again an absorptive grating.
If the molecular beam were spatially coherent, right from the source, the first grating could also be a phase grating with 100\% transmission. This has been demonstrated for cold atoms~\cite{Cahn1997b} and Bose-Einstein condensates~\cite{Denschlag2000}. However, the preparation of coherent macromolecular beams has remained a challenge throughout the years and even though there is good progress in the development of various source types, coherence still needs to be enhanced, by virtue of spatial selection. 

In Fig.~\ref{fig:kdtli} we show a sketch of the near-field Kapitza-Dirac-Talbot-Lau interferometer (KDTLI)~\cite{Gerlich2007c}. Its mechanical gratings G1 and G3 are etched into a 160~nm thick nanomechanical SiN$_x$ membrane with a period of $d=266$~nm and open slits as narrow as $s=110$~nm. While the width of each individual opening is only defined by the etching process to within 2-5~nm accuracy, the average grating period is defined to better than 50~ppm.  

The first grating G1 represents an array of 'nano-illuminators'~\cite{Jahns1979}. Molecules impinging onto G1 under any angle of incidence will have their transverse wave function restricted and therefore their momentum coherently spread, in accordance with Heisenberg's uncertainty relation. We can also see it as the preparation of an approximate cylindrical Huygens wavelet, the wave front of which is well defined because the point of origin is so small.
It was E. Lau~\cite{Lau1948} who realized that spatially coherent illumination of the second grating, at least over two neighboring slits, can be achieved if the distance $L$ between G1 and G2 is chosen close to the Talbot length $L_T=d^2/\lambda_\mathrm{dB}$.
Diffraction at the central optical grating G2 then leads to the lens-less formation of a grating self-image, which can only be explained by near-field wave interference. The transmitted molecules thus form a particle density pattern in free space, with a characteristic length scale of $L\simeq L_T$ behind G2, which can be captured and imaged on a screen~\cite{Juffmann2009b}.

\begin{figure}[htb]
	\centering
	\includegraphics[width=0.9\linewidth]{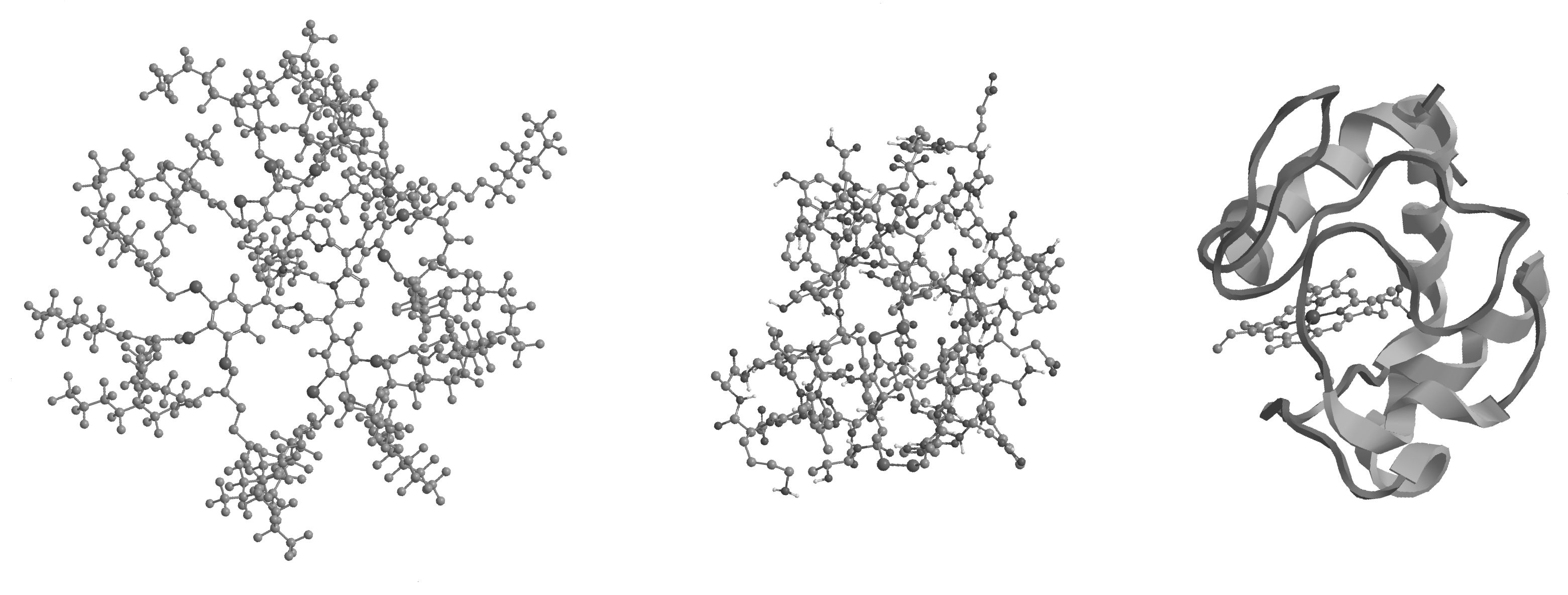}
	\caption{The functionalized porphyrin TPPF20 (left) is the largest object for which matter-wave interference has been observed so far. It compares in complexity and mass with insulin (middle) or cytochrome C (right). The extension of TPPF20 can reach up to 50~\AA.}
	\label{fig:kdtlicompare}
\end{figure}

It is, however, often favorable to use a third mask, G3, whose period equals that of the emerging molecular fringes. The interference pattern is then encoded in the number of molecules transmitted through G3 as a function of the transverse position of G3. Recording the interference pattern by measuring the molecular count rate as a function of transverse position of G3 has become the standard technique in Kapitza-Dirac-Talbot-Lau interferometry and has allowed to demonstrate matter-wave coherence with the most massive particles to date (see Fig.~\ref{fig:kdtlicompare}).
This achievement became possible because of the highly developed capabilities in chemical synthesis in the group of Marcel Mayor at the University of Basel. They provided porphyrin derivatives where numerous peripheral hydrogen atoms were substituted by massive perfluoroalkyl chains. This trick allows increasing the particle mass, maintaining a low polarizability-to-mass ratio and therefore achieving a good vapor pressure in a sublimation or vaporization process~\cite{Tuxen2011}. We measured high-contrast matter-wave interference with these particles, even though they are fragile and $\sim$500~K hot \cite{Eibenberger2013}. 

These functionalized molecules are already in the complexity class of proteins and we compare them pictorially to the structure of the proteins insulin and cytochrome C in Fig.~\ref{fig:kdtlicompare}. TPPF20 (C$_{284}$H$_{190}$F$_{320}$N$_{4}$S$_{12}$) is made from 810 covalently bound atoms with a total molecular mass of 10\,123~amu and is the most massive object for which matter-wave interference has been observed so far~\cite{Eibenberger2013}. Insulin, for example, is composed of 777 atoms in 51 amino acids (C$_{254}$H$_{377}$N$_{65}$O$_{75}$S$_6$) with a total mass of about 5\,743~amu, and cytochrome C is composed of 104 amino acids, 1\,776 atoms and has a total mass of 12\,430~amu (C$_{541}$H$_{910}$FeN$_{145}$O$_{175}$S$_{4}$). Interference with such proteins is still work in preparation, since even for the same interferometer concept and scale as before, the source preparation and molecule detection are substantially more demanding for large biomolecules.  

\paragraph{The role of absorption during the interaction with the optical grating}

So far we have described an idealized phase grating in G2 and neglected absorption. However, when the optical power or the molecular absorption cross section are high, photons can be absorbed in the standing light wave. The probability of photon absorption is highest in the anti-nodes and lowest in the nodes of the standing light wave. The mean number of absorbed photons $n_0$ during the transit through light grating formed by a Gaussian laser beam depends on its waist $w_y$, the molecular velocity through the grating $v_z$, and the wavelength of the laser $\lambda_L$, 

\begin{equation}\label{absorption}
n(x)=n_0 \cdot \cos^2(k_L x) = \frac{8 \sigma_{abs} \lambda_\mathrm{L} P}{\sqrt{2\pi}hc w_y v_z}\cdot \cos^2(k_L x).
\end{equation}

The absorption of a photon is accompanied by a momentum transfer of $\Delta p=\hbar k_L$ to the molecule. If the molecules were exposed to incoherent thermal light they would 'know' whether the photon came from the left or the right and receive a corresponding momentum kick. Inside the coherent standing laser light field, however, the photon itself is in a superposition of 'coming from the left' and 'coming from the right'. Upon absorption, the molecular matter-wave thus also ends up of in a superposition of moving to the left \textit{and} to the right. 

In far-field diffraction we cannot distinguish between those two possibilities since both statistical and coherent absorption populate the same momentum states. Inside the KDTLI interferometer, however, the situation is different: the wave function of a single molecule may pass two neighboring anti-nodes of G2 in momentum states that would not overlap at $G3$. The additional recoil may however refocus them and thus enable their interference. 

The observed process is also intriguing with regard to measurement and the possibility of entanglement. Absorption labels a molecule by its internal energy: every additional 532~nm photon raises the internal temperature of a C$_{70}$ molecule by about 139~K~\cite{Hornberger2005}.
This creates a correlation between the internal and the center-of-mass state -- very much like in a Ramsey-Bord\'e beam splitter \cite{Borde1989}. It differs, however, in that there is no simple Rabi cycling in macromolecules to reverse the absorption process coherently. 

In order to interfere with itself the molecule needs to remain in the same internal state along all paths. Upon absorption the molecule loses its ability to interfere with all parts of the superposition state which have not changed their internal energy. Absorption thus transfers each delocalized molecule into a mixture of temperature states, depending on the number of photons they absorb, and on the instant they do. However, in spite of all complexity, the internal dynamics remains unitary within each temperature class. The quantum random walk in momentum space is thus accompanied by a division of the molecular coherence into different temperature classes. Within each internal energy class, de Broglie coherence is maintained.
\begin{figure} [thb]
	\centering
	\includegraphics[width=\linewidth]{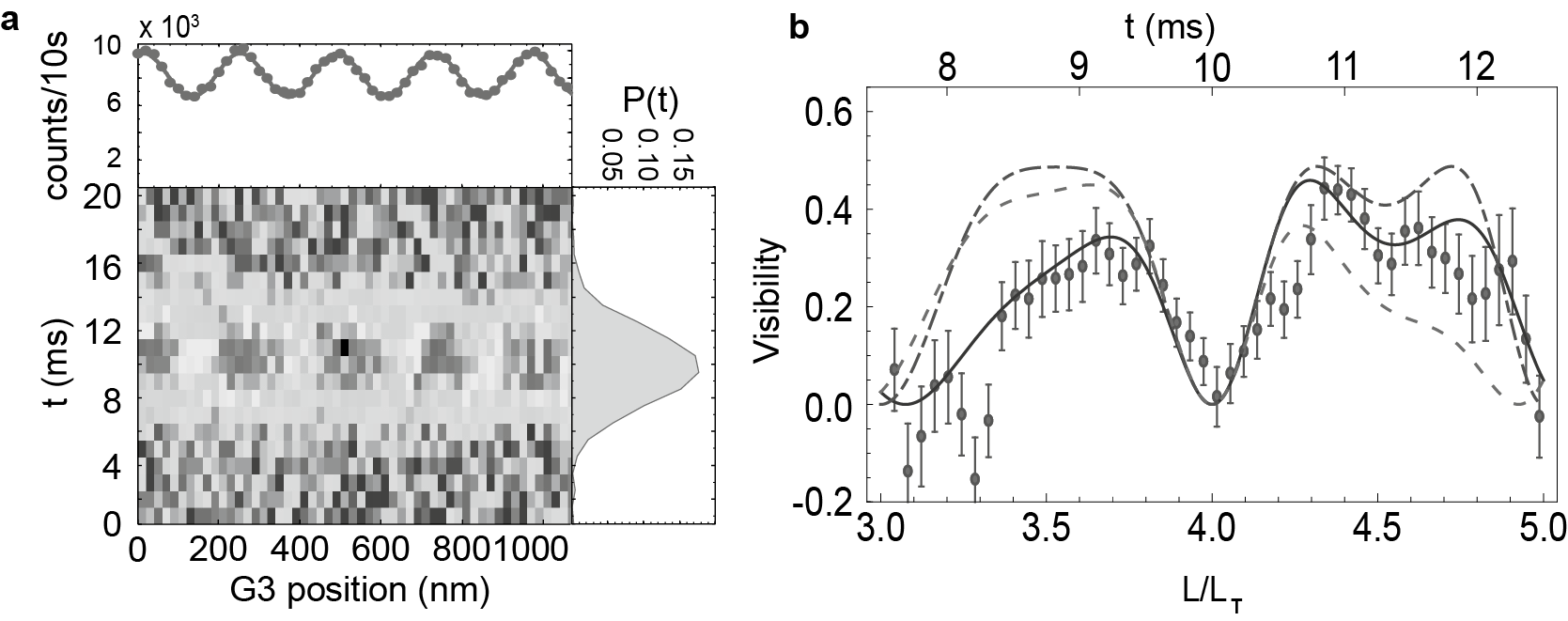}
	\caption{Coherent absorption in KDTLI: a) The time-of-flight resolved molecular interference pattern has an almost Gaussian velocity distribution (right) and a near-sinusoidal integrated molecular density distribution in real space, as also expected in theory. b) Interference fringe visibility as a function of the scaled interferometer length $L/L_T=L m v/h d^2$. Both a pure phase grating model (dash-dotted) and a model assuming a random walk in momentum space (dashed) fail to reproduce the experiment. The full quantum model (continuous line) accounts for coherent single-photon absorption, periodic phase shifts, as well as a measurement-induced grating (see text).}
	\label{fig:GratingRecoil}
\end{figure}
Absorptive heating in the light wave realizes a special case of a photo-depletion grating (see below), which 'depletes' a given molecular temperature class with spatially periodic grating structure. Similar to diffraction at a mechanical grating, where molecules 'know' about the nearby grating bars even when they are transmitted by the open slits, the unheated molecules 'know' about the possibility of being heated by the photons in the standing light wave. Interestingly, and different from the situation at nanomechanical gratings, the 'depleted' states can form their own interference class: all spatial wave functions with the same number of photons absorbed in different places at the same time are still de Broglie-coherent. 

A proper model has to take into account all aspects of the molecule-grating interaction: the phase shift imparted by virtue of the dipole force, the effects of coherent absorption and even the measurement-induced grating. These contributions were analyzed in detail by our theory partners around Klaus Hornberger at the University of Duisburg-Essen. 

In Fig.~\ref{fig:GratingRecoil} we compare their model with our time-of-flight resolved experiments with C$_{70}$ molecules. Different molecular velocities correspond to different Talbot lengths $L_T$. In the KDTLI, with a distance $L$ between consecutive gratings, the experimentally accessible velocities for C$_{70}$ span over several Talbot orders $L/L_T$, allowing us to study the interference visibility as a function of Talbot order. The dashed-dotted line in Fig.~\ref{fig:GratingRecoil} compares our data with the assumption of a pure phase grating, the dashed line represents a model assuming incoherent stochastic absorption, and the solid line represents a full quantum model including coherent absorption in an optical grating~\cite{Cotter2015}. The agreement with the experiment is clearly best for the complete model. 

One may wonder, why decoherence by photon emission, as observed for atoms~\cite{Chapman1995a,Kokorowski2001} and fullerenes~\cite{Hackermuller2004,Hornberger2005}, does not destroy the interferogram. In many complex molecules this process can be suppressed, because of vibrational relaxation on the picosecond scale and inter-system crossing (singlet-triplet) on the nanoscecond scale~\cite{Dresselhaus1998}. This eliminates the stochastic recoil of the spontaneous emission process and the distribution of which-path information to the environment. Fluorescence or phosphorescence are not entirely excluded for all molecules, and might localize the molecule to half of the wavelength of the emitted radiation. This has been taken into account in a recent theoretical model, too~\cite{Walter2016a}.

\subsubsection{Photo-depletion gratings}
Measurement-induced gratings are not only implicitly present in KDTL interferometry, but have been also explicitly used in atom optics, before. Diffraction at depopulation gratings was for instance observed with argon atoms that were optically pumped from a detectable metastable state to the undetected ground state~\cite{Abfalterer1997c}. Pulsed optical gratings in the time-domain were also essential in realizing a complete atom interferometer~\cite{Fray2004}.

The biggest difference between quantum optics of atoms and many-body systems consists in the way we can address the latter.
Resonant interactions are typically precluded, but photon absorption can still deplete particle beams in various ways.

Single-photon ionization gratings were suggested to enable new tests of high-mass quantum physics in metal cluster interferometry ~\cite{Reiger2006a}. The idea was theoretically corroborated~\cite{Nimmrichter2011b}, and experimentally demonstrated~\cite{Haslinger2013b} in OTIMA interferometry, a matter-wave interferometer with three optical photodepletion gratings in the time-domain. Vacuum ultraviolet nanosecond laser pulses allowed establishing the probably shortest conceivable optical grating period (with $d=78.5$~nm) and to provide a photon energy high enough to ionize all metal clusters, many dielectric particles, and many tryptophan-rich peptides.
Single-photon ionization is independent of any specific internal resonance and therefore counts as an ingredient for a 'universal' diffraction element in molecular quantum optics. 

Even particles with ionization energies exceeding the energy of a single photon can often be treated using a similar concept. 
Weakly bound van der Waals clusters may absorb a photon and rather dissociate than ionize - and thus also be removed from the particle beam.
Diffraction gratings based on that effect have recently been demonstrated for hexafluorobenzene and vanillin~\cite{Dorre2014h}.

\section{Quantum enhanced measurements} 
The sensitivity of quantum phases to external perturbations is friend and foe at the same time. 
While it increases the demand for ever better experimental skills, it also creates new opportunities for
refined measurements on individualized particles in ultra-high vacuum. Chemistry, biology, pharmacy, and medicine depend on detailed knowledge of molecular properties, many of which may be obtained through interactions with static or dynamic electrical or magnetic fields, with radiation at all wavelengths or collisions with controlled probe particles. 

Matter-wave interferometry can contribute to the field by generating free-flying nanoscale structured molecular density patterns that will shift in response to the external forces. The physics behind this idea is similar to that of atom interferometric sensors, which are typically geared towards measuring inertial forces, caused by gravity and rotations~\cite{Cronin2009,Tino2014}. 

Since the spatial resolution of beam shifts in matter-wave experiments can exceed that of classical deflectometers by far, the quantum advantage bears great potential for molecule metrology, too. This is best exploited if interferometers can handle a large range of diverse atoms, molecules, clusters, covering several orders of magnitude in mass, different internal configurations and excitations. This is what KDTLI and OTIMA interferometry can offer. 

Over the past years many different particle properties have been studied in Vienna. This comprises measurements of the static \cite{Berninger2007b} and the optical \cite{Hackermuller2007a,Hornberger2009a} polarizability, permanent \cite{Eibenberger2011}, and vibration-induced electric dipole moments~\cite{Gring2010b}. 
Molecule interferometry was shown to complement mass spectrometry in discriminating conformational isomers -- which only differ in their atomic arrangement, but not their mass~\cite{Tuxen2010b} -- and to attribute the origin of molecular fragmentation~\cite{Gerlich2008a}.
Here we focus on a recent experiment demonstrating the potential of matter-wave assisted spectroscopy on molecules \cite{Nimmrichter2008,Eibenberger2014}. The idea finds again a close analogy in atom optics, where a photon recoil was used to determine $h/m_\text{Cs}$ and from this the fine structure constant with high precision~\cite{Bouchendira2011}. 

Our interest is directed towards the absolute absorption cross section $\sigma_{abs}$ which has been notoriously difficult to assess for large organic non-fluorescent molecules in dilute beams. There are multiple challenges in classical molecular beam physics: Firstly, many large organic species have low vapor pressure, typically smaller than 0.1~mbar, even close to their decomposition temperature. If the absorption cross section is as small as $\sigma_{abs}=10^{-17}$~cm$^{-2}$ a thermal vapor would need to be extended over 40~cm to absorb the transmitted fraction to $1/e$ of the incident light. But already in a unperturbed molecular beam, extracted from this vapor and after free-flight over about 2~m distance, the density has dropped to about $10^{10}$~cm$^{-3}$ where direct absorption would require a prohibitively long vapor cell. Very dilute molecular beams can often still be observed in fluorescence or using an 'action mechanism' such as the evaporation of an adsorbed noble gas atom~\cite{Fielicke2004a}. But this does not apply to fragile cluster aggregates, non-fluorescent biomolecules, or many molecules of chemical or even astrophysical interest.
\begin{figure} [htb]
	\centering
	\includegraphics[width=0.6\linewidth]{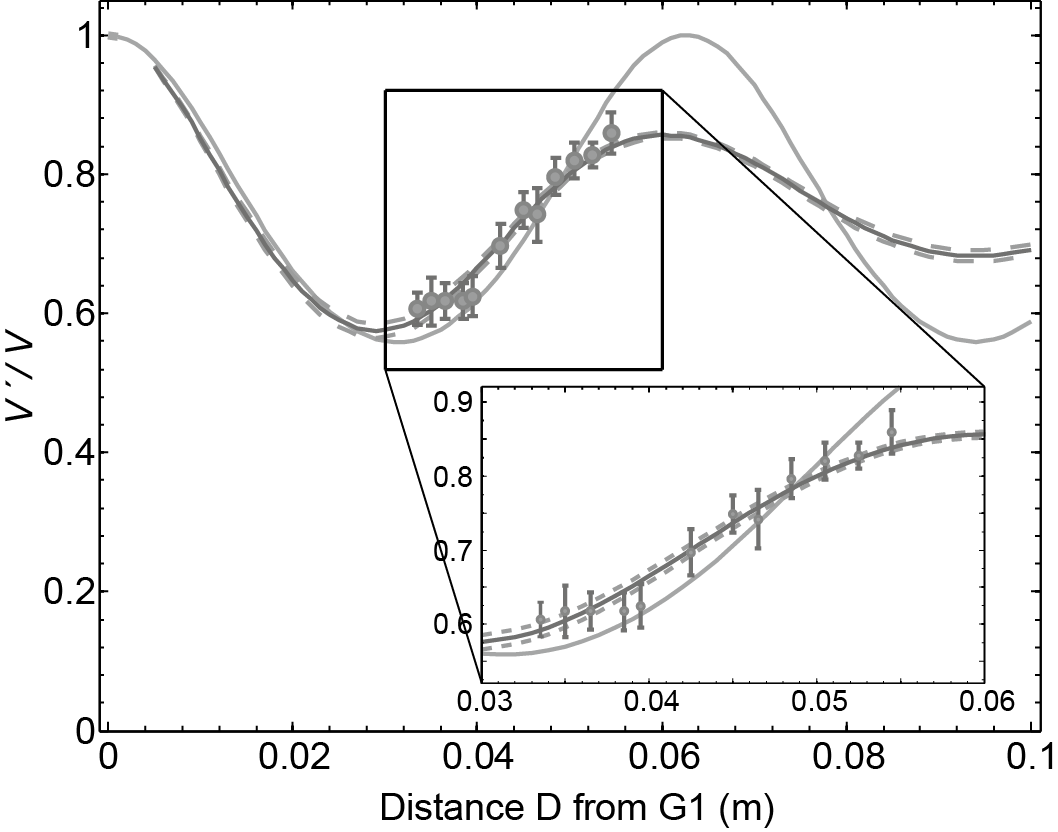}
	\caption{The visibility reduction factor $R=V'/V$ depends on the location where the laser hits the molecule relative to the inter-grating separation. The perturbation is strongest close to G2. A fit of the data to the reduced visibility allows extracting the absolute cross section with high resolution.}
	\label{fig:KDTLILaser}
\end{figure}
Starting from the experiment, sketched in Fig.~\ref{fig:kdtli}, we realize that the wave function associated with every individual molecules is first delocalized, diffracted and then recombined to obtain the interferogram that is probed by scanning grating G3. In KDTLI, the molecular nanostructure has a periodicity of 266~nm, which can be resolved with a position accuracy of about 10~nm for the signal-to-noise that is typical for molecules below 1000~amu. We now add a \textit{running} laser beam at distance $D$ from G1 along the grating $k$-vector which is sufficiently intense to ensure the absorption of 0.1-0.3 photons per molecule on average. A delocalized molecule that absorbs a photon will experience a recoil by the photon momentum $\Delta p=\hbar k$ and be excited to a higher-lying electronic state. It exhibits an interferogram displaced by the distance $s=\lambda_{\mathrm{dB}}D/\lambda_K$, where $\lambda_K$ is the wavelength of the running laser beam. In our particular example the photon wavelength of 532~nm was comparable to the wave packet separation close to the second grating. And yet, high-contrast de Broglie interference can be maintained since the absorption process is phase coherent. 
The total pattern is the sum of all shifted and unshifted ones, but there are only two discrete options. The overall contrast is thus reduced, but by an amount which can be unambiguously correlated with the number of absorbed photons. When the laser beam parameters are well known, one can extract the absolute absorption cross section with few percent precision. 
Photon absorption diminishes the unperturbed interference fringe visibility $V$ and leads to a reduced visibility $V^{\prime}= \langle R \rangle_{v} V$, with the velocity averaged reduction factor $\langle R\rangle _v$~\cite{Eibenberger2014} 

\begin{equation}
\langle R\rangle _v=\left |\int_0^{^\infty} dv_z P(v_z) \exp\left(-n_0 \left[1-\exp(2\pi i s/d)\right]  \right) \right|.
\label{eqn:Vreduction}
\end{equation}

The realization of this idea allowed characterizing the absolute cross section for the fullerenes $C_{70}$ to within 2\%. For this proof-of-principle experiment a high-power continuous solid state laser was used. Future upgrades can be based on widely tunable radiation as provided by a dye laser, titanium sapphire laser and their higher harmonics. The scheme is also open to a number of multi-photon combinations~\cite{Rodewald2016}. 

\section{Molecular beam sources for nanoscale organic matter and biomolecules}
If we want to pursue quantum experiments with high mass objects in analogy to advanced atom interferometry, it is necessary to prepare beams of isolated neutral particles that are sufficiently slow and cold, selected in mass, and to some extent also geometry. 
The neutrality requirement is not fundamental -- as a matter of fact electron interferometry is an integral part of electron microscopy and holography~\cite{Tonomura1999} -- however it is of practical importance, since charged particles are easily perturbed by external fields. At present, the helium cation He$^+$ is still the most massive species in fundamental matter-wave physics with charged particles~\cite{Hasselbach2010}. Other than that, we require the particle momentum to be smaller than $10^7$~amu~$\times$~m/s, corresponding to a minimum de Broglie wavelength of $\lambda_\mathrm{dB}=40$~fm. This is the target value for the next-generation of macromolecule interferometers in Vienna.

In single-grating far-field diffraction, mass selection is important, since the scattering angle scales inversely with particle mass. In interferometric near-field self-imaging, however, the period of the fringe pattern is always determined by the interferometer geometry. The particle mass then determines the fringe contrast, via the match or mismatch between the Talbot length $L_\mathrm{T}=d^2/\lambda_\mathrm{dB}$ or Talbot time $L_\mathrm{T}=d^2 m/h$ and multiples of the grating distance. In practice, advanced near-field interferometers can tolerate a span of $\Delta \lambda_\mathrm{dB}/\lambda_\mathrm{dB} \simeq 10\%$. Deviations from that rule of thumb emerge for $m>10^5$~amu, because of dispersive phase shifts.  

A key difference between molecule and atom interferometry is the internal particle structure. In our current experiments, we are mostly interested in \textit{de Broglie }interference, that is the evolution of the particle's center-of-mass motion. However, we recall that an $N-$body system can store energy in 3 translational, 3 rotational and $3N-6$ vibrational modes, most of which will be (highly) excited at room temperature. 

Vibrational modes, with an energy of $0.01-0.3$~eV, may be frozen out in the presence of cryogenically cold buffer gas. However, even then a large molecule may be trapped in a local energy minimum of a vast conformational landscape.

It is even more challenging to approach the rotational ground state. An insulin molecule, for example, with a mass of 5\,808~amu, is composed of 51 amino acids with a total moment of inertia of about $9\times 10^{-42}$~kg~m$^2$.
Its rotational energy levels therefore scale like $E_{\mathrm{rot}}\simeq 40~\mu$K$\times J(J+1)$ and are excited to an average rotational quantum number around a few thousand at room temperature and still around $J\simeq 100$ in a superfluid helium droplet at $T=380$~mK~\cite{Toennies2001}. Human hemoglobin ($m=68\,600$~amu), containing about 4\,800 atoms in a toroidal diameter of about 60~\AA\,  has a moment of inertia of $4 \times 10^{-40}$~kg~m$^2$, pushing the temperature requirements for rotational ground state cooling by another factor of ten.  

When we further consider all electronic, fine, hyperfine, and Zeeman sublevels, it is safe to say that all molecules will populate different statistical combinations of their internal states and every particle will only interfere with itself, never with another one. High-contrast de Broglie interferences is still possible as long as the internal states do not provide information about the molecular position.

\subsection{Thermal beams and thermal molecules in supersonic beams}
Thermal sublimation from the solid phase or evaporation from the liquid phase works well for many molecules up to about 1\,000~amu. This includes all nucleobases, various porphyrins (600-800~amu), the fullerenes C$_{60}$ (720\,amu) and C$_{70}$ (840\,amu) but also vitamins, such as beta-carotene (pre-vitamin A) or vitamin E ($\alpha$-tocopherol). Most of them have been studied in KDTL interferometry. 

Vanillin and caffeine have sufficient thermal vapor pressure to form a thermal molecular beam that can be clustered into complex aggregates, when injected into an adiabatically expanding supersonic noble gas jet. This has been successfully used in OTIMA interferometry~\cite{Dorre2014h}. Since they are bound by van der Waals forces, these cluster are prone to dissociation after photon absorption. This makes them susceptible for photo-depletion gratings and photodepletion spectroscopy, in general.

\subsection{Tailoring organic materials for improved volatility}
The required sublimation or evaporation temperature can be reduced by chemical functionalization. Perfluoroalkyl chains such as S$($CH$_{2})_{2}$C$_{8}$F$_{17}$ have been found to substantially reduce the intramolecular binding~\cite{Tuxen2011}, since the high electro-negativity of the fluorine atoms binds the charge and reduces the electrical  polarizability. Such functionalization has allowed to volatilize porphyrin derivatives for successful quantum interference experiments beyond $10^4$~amu~\cite{Eibenberger2013}. The same idea has enabled thermal beams of oligopeptides, too~\cite{Schatti2017}. Even though one might think that such modifications are artificial, it turns out that medicine and pharmacy are full of fluorinated compounds~\cite{Purser2008}. This suggests that also quantum-interference enhanced measurements on \textit{tailored} biomolecules can provide relevant data. 

\subsection{Laser injection of large peptides into expanding noble gases}
Many biomolecules will denature or fragment abundantly, when they are heated for extended periods of time.
This can be prevented by reducing the interaction time: Focusing a nanosecond laser beam  onto an area of 1~mm$^{2}$ with mJ pulse energy can heat a sample at a rate of $\sim10^{11}$~K/s. Even fragile molecules may thus leave the surface before too much energy is absorbed. In order to prevent post-desorption fragmentation the particles are entrained in an adiabatically expanding noble gas jet, released by a nearby short-pulse nozzle. At stagnation pressures of several bars and gas pulse times around 20-30~$\mu$s the number of collisions is high enough to cool even large peptides. While the current limits of this technique are still being explored, it has been shown that even polypeptides composed of more than a dozen amino acids could successfully be brought into the gas phase using this technique~\cite{Geyer2016b}.
\begin{figure} [htb]
	\centering
	\includegraphics[width=0.8\linewidth]{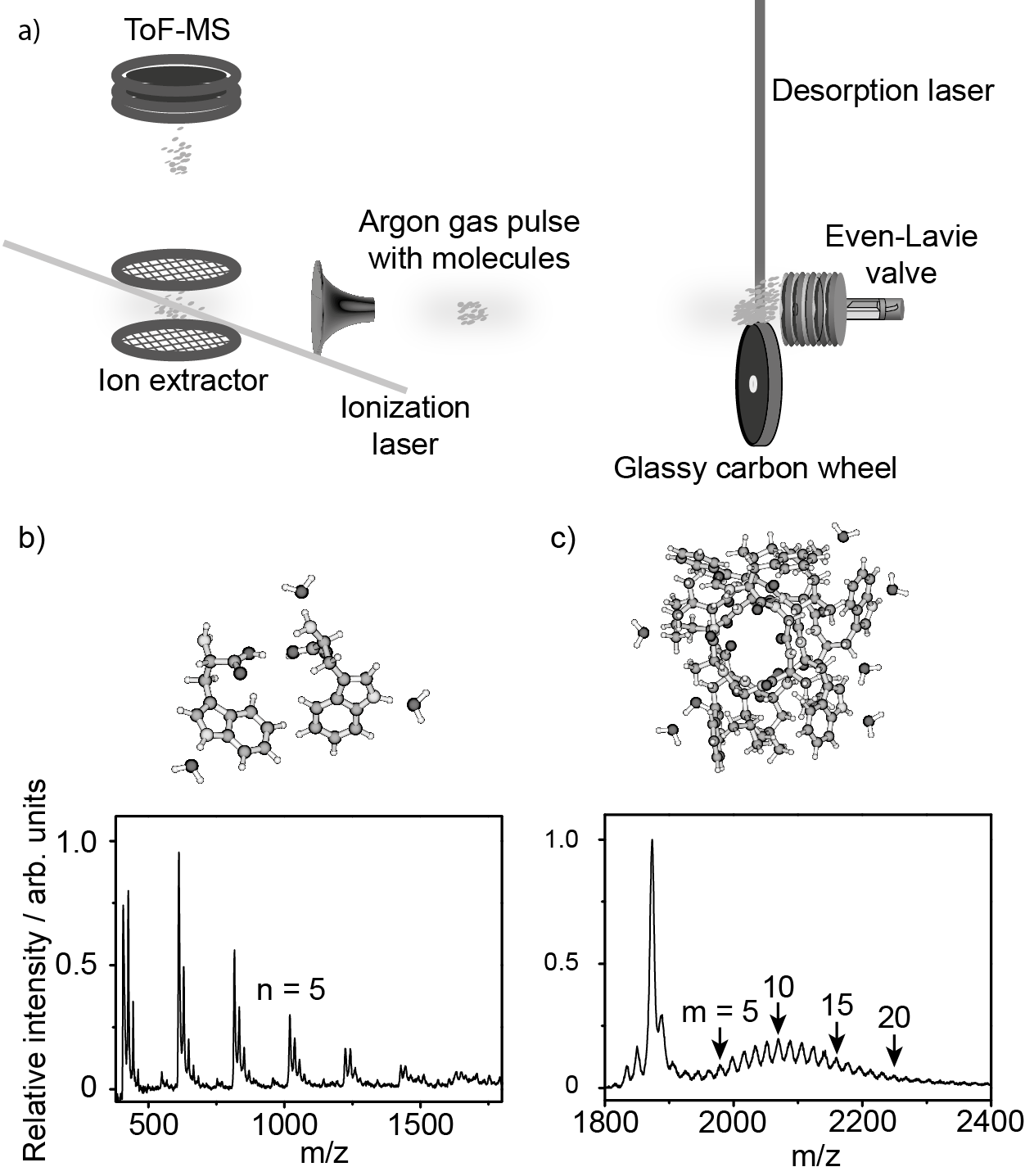}
	\caption{a) The pulsed nanosecond laser light desorbs molecules from a glassy carbon wheel which are immediately entrained into an expanding noble gas jet. The molecules are post-ionized with vacuum ultraviolet light (157~nm) and detected in a TOF-MS. This source generates hydrated cluster beams of tryptophan (b) as well as of polypeptides like gramicidin (c) or indolicidin. Here, $n$ corresponds to the number of clustered amino acids and $m$ to the number of attached water molecules.}
	\label{fig:Supersonic}
\end{figure}
We can also proceed one step further and ask for the influence of microhydration, i.e. the controlled addition of water molecules to the free biomolecule. Such studies shall identify the role of the native environment on the molecular properties and the possible role of evaporation as an enabling agent in optical diffraction gratings or a cause of evaporative decoherence~\cite{Geyer2016b}.

The implementation of such as source is shown in Fig.~\ref{fig:Supersonic}a): Biomolecular powder is picked up by a felt wheel and pressed onto a glassy carbon wheel~\cite{Gahlmann2008}. The thin biomolecular layer is then desorbed by a nanosecond laser pulse. The expanding noble gas jet entrains and cools the molecules before they can dissociate. This allows launching amino acid clusters (Fig.~\ref{fig:Supersonic}b), and peptides (Fig.~\ref{fig:Supersonic}c), also with a micro-hydration shell. 

\subsection{Laser-induced acoustic desorption}
We may also ask whether it is possible to launch particles from a surface in vacuum without a direct exposure to the laser. This question has been explored by several groups in physical chemistry~\cite{Zinovev2011a} and \textit{Laser-Induced Acoustic Desorption} (LIAD) has been successfully used to even load lowly charged individual viruses and bacterial cells into an ion trap~\cite{Peng2004}.
\begin{figure} [htb]
	\centering
	\includegraphics[width=0.9\linewidth]{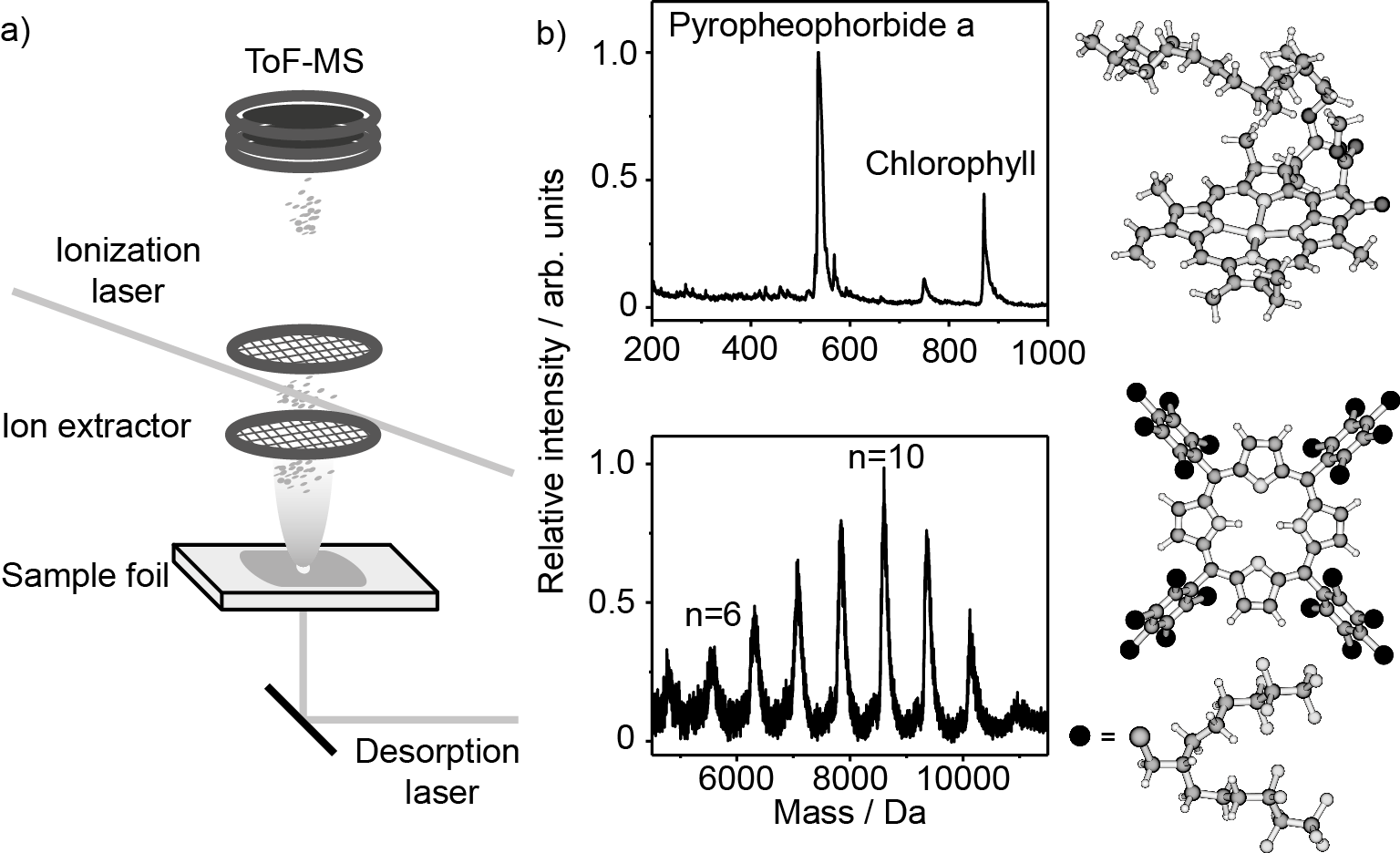}
	\caption{a) An energetic nanosecond laser pulse at 355~nm hits the back-side of a thin titanium or tantalum foil. The ablation shock releases biomolecules from the front side of the metal sheet. The molecular velocity averages around 160~m/s for molecules in the mass range of porphyrins desorbed from tantalum. Surprisingly, we achieve velocities as low as 30~m/s when launching molecules from titanium foil. The internal molecular temperature has not been determined. For a given experimental setup the molecule velocity scales like $v\propto m^{-1/2}$. b) The mass spectrum of chlorophyll shows that the fragile parent molecule is successfully released from the surface and detected after photo-ionization at 157~nm. The TPP derivatives represent an entire library of functionalized biochromophores which fly intact and slow using LIAD and can also be photo-ionized. Here, $n$ is the number of attached side chains.} 
	\label{fig:LIAD}
\end{figure}
Our goal here is to prepare a beam of macromolecules, sufficiently intense and slow to comply with the coherence conditions in molecule interferometry. We performed demonstration experiments using the setup shown in Fig.~\ref{fig:LIAD}.
A nanosecond laser pulse (1~mJ) is directed onto the backside of a thin metal sheet. This sheet is only 10-20~$\mu$m thick and covered on its  front-side by an even thinner layer of molecules. When the laser beam hits the backside of the foil, it generates a fast metal plume which imparts a strong recoil onto the sheet and also releases molecules from the front side.

In our experiment (see Fig.~\ref{fig:LIAD}) we were able to prepare slow beams of neutral biomolecules and functionalized organic libraries~\cite{Sezer2015b}, which were post-ionized using vacuum ultraviolet light at 157~nm and mass-analyzed in time-of-flight mass spectroscopy. We observed neutral beams of chlorophyll as well as large functionalized porphyrins.  

While intact chlorophyll has an average speed of 50 to 160~m/s, depending on the foil material, tailored porphyrins with $m=10^4$~amu were measured to have a mean velocity around $v=20$~m/s. Their kinetic energy of 20~meV corresponds to $T\simeq 270^{\circ}$\,C assuming a thermal launching process. This is supported by the fact that the measured velocity scales like $v\propto m^{-1/2}$. Although the LIAD process is still not entirely understood it seems to have both a thermal and a mechanical contribution~\cite{Zinovev2011a,Shea2007}. 

We also discovered that LIAD can produce nanoparticles from pristine silicon wafers \cite{Asenbaum2013} and nanostructured silicon templates with tailored nanorods~\cite{Kuhn2016}. Particles as massive as $10^{10}$~amu were broken off and launched at velocities between 0.2-30~m/s. The kinetic energy of such a particle flying at 10~m/s amounts to 5.2~keV. It seems that during the launching process energy stored in the deformation of the substrate is released. Even starting from such high energy, cooling in a high finesse cavity was observed to reduce the particle's transverse kinetic energy by about a factor of thirty~\cite{Asenbaum2013}. 

This observation was also a first step in a series of experiments towards preparing particles for quantum interferometry with masses around $10^6-10^{7}$~amu. This ongoing research work complements efforts in other places where active feedback cooling~\cite{Gieseler2012} or cavity cooling~\cite{Kiesel2013,Millen2014g} of trapped and neutral particles is also pursued.

\subsection{Molecular ions as the basis for neutral molecular beams}
Since the advent of \textit{Matrix Assisted Laser Desorption Ionization (MALDI)} \cite{Tanaka1988a,Karas1989} and \textit{Electrospray Ionization (ESI)}~\cite{Fenn1989} charged protein and DNA beams have been heavily used in mass spectrometry. 
MALDI differs from laser desorption (LD) in that the analyte molecules (proteins, DNA, etc.) are embedded in an acidic molecular matrix on a substrate. The matrix molecules are optimized for high absorption around 330-350~nm which can be excited by the radiation of a nitrogen laser or frequency tripled Nd:YAG laser. Their wavelength is red-shifted with respect to the absorption spectrum of peptide bonds ($\sim 213$~nm), the aromatic amino acids, and nucleobases (260-290~nm). The photon energy is therefore dominantly deposited in the matrix molecules which evaporate and entrain the analyte particles. Since the matrix consists of proton donors it can also protonate and charge the analyte molecules.
MALDI typically launches singly charged biomolecules of either polarity in high vacuum, at moderate repetition rate and well suited for short experiments with small samples. In contrast to that, electrospray ionization is better suited for preparing steady continuous beams.
\begin{figure}
	\centering
	\includegraphics[width=0.75\linewidth]{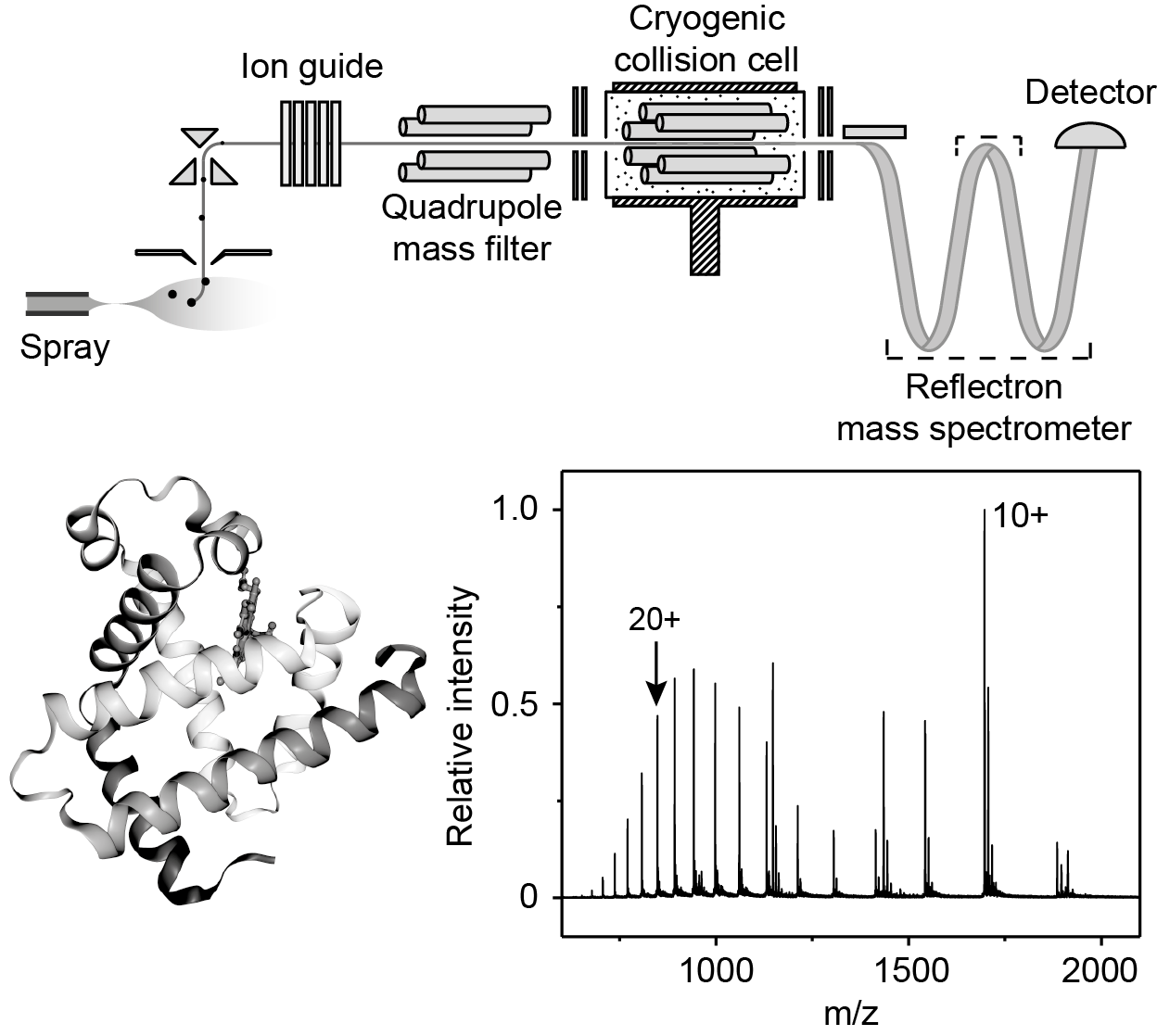}
	\caption{Electrospray ionization yields highly charged biomolecules which are mass-selected and subsequently cooled in the presence of a cryogenic gas. The detection takes place in a high-resolution time-of-flight mass spectrometer. For proteins the size of myoglobin (153 amino acids, 17\,000 amu), ESI natively ejects highly charged molecules. Collisions with bi-polar air can reduce this to a single charge per molecule.}
	\label{fig:ESI}
\end{figure}
In electrospray ionization analyte molecules are filled into a narrow capillary, which is placed at a distance to an opening in the vacuum machine. At high electric fields, the liquid forms a cone which breaks into a filament and further into droplets which partially dry in air and also eject further droplets, because of the growing imbalance between Coulomb repulsion and surface tension. A series of emission and evaporation processes leads finally to the release of individual, unsolvated analyte molecules. Hence, the mass-to-charge ratio in polypeptides and proteins usually ranges between 1\,000-2\,000~amu/e in ESI experiments. This is ideal for mass analysis in commercial quadrupole mass filters, but less advantageous for quantum experiments with neutral beams.
Charge exchange techniques from aerosol science can reduce the charge per particle by collisions in bi-polar air \cite{Bacher2001} before they are transferred into high vacuum through ion guides and a differential pumping system. The molecules are mass-selected in a quadrupole mass filter in high vacuum. A subsequent hexapole guide allows collisions with cryogenically cold (typically 60~K) neutral buffer gas (see Fig.~\ref{fig:ESI}). 

ESI works over a very wide mass scale, from amino acids to viruses. However, the manipulation, control and mass spectrometry needs to be adapted for a specific mass scale. The setup in Vienna is currently optimized for a maximal mass-to-charge ratio of up to 30\,000~amu/e. The extraction of neutral particles by photo-depletion processes is an open challenge still under investigation~\cite{Sezer2017}.

\section{Conclusion}
Quantum optics with complex molecules has many far-reaching goals and faces a plethora of intriguing challenges. The field shares many concepts and solutions with atom interferometry and atom optics, however, enriched by the complexity of strongly bound systems, composed of 1\,000 atoms at present and possibly 100\,000 atoms in the future. 

The branch of organic and biomolecular interferometry has been developing with new ideas for novel quantum and decoherence experiments, but also new hope for a radically different way of measuring biomolecular matter under controlled conditions. These experiments aim at single-molecule sensitivity, either isolated from the surroundings or hydrated with a controlled number of water molecules and also in interaction with single photons.

Matter-wave physics with the building blocks of life is still a young research activity which profits from vigorous and fruitful interactions with researchers in physical chemistry, computational and biochemistry, nanoimaging and nanotechnology. Many quantum manipulation and interference techniques have already been developed. Further success is strongly tied with rapid development in molecular beam technologies.  

\section{Acknowledgements}

This summary encompasses a selection of years of fruitful research in the Quantum Nanophysics Group in Vienna, including present and former members, as well as our collaboration partners around Ori Cheshnovsky (Tel Aviv University, nanofabrication and optical imaging), Klaus Hornberger, Benjamin Stickler, and Stefan Nimmrichter (University of Duisburg-Essen, Quantum theory), Stefan Scheel (University of Rostock, Casimir-Polder theory), and Marcel Mayor and Valentin K\"ohler (University of Basel, chemical synthesis and functionalization). We thank them all for their inspiring and important input over many years that is only partially reflected in this lecture. MA is grateful to Anton Zeilinger (University of Vienna) for many years of fruitful collaboration on macromolecule interferometry. We thank in particular Stefan Gerlich, Lukas Mairhofer, and Joe Cotter for their contributions to KDTLI interferometry and Maxime Debiossac and Moritz Kriegleder for recording the mass spectrum of myoglobin. We are grateful to financial support through the European Research Council project 320694, the European Commission within project 304886 as well as the Austrian Science Funds project W1210-3.

\end{document}